\documentclass[twocolumn]{aastex62}

\usepackage[utf8]{inputenc}
\DeclareUnicodeCharacter{2212}{-}
\usepackage{graphicx}
\graphicspath{{./}{./figures/}{./tex/figures/}}
\usepackage{amsmath}
\usepackage{afterpage}
\usepackage{enumitem}
\usepackage{xcolor}
\setenumerate{itemsep=0mm}

\usepackage{lineno}

\definecolor{newblue}{cmyk}{1,0.7,0,0}

\newcommand{\addref}[1]{\textcolor{red}{[ADD REFERENCE] }}

\usepackage{soul} 
\usepackage{amsmath}
\usepackage{amssymb}
\usepackage{xspace}
\usepackage{xifthen}
\usepackage{eso-pic}

\newcommand{\Gaia}{{\it Gaia}\xspace}

\definecolor{forestgreen}{HTML}{228B22}
\definecolor{urlblue}{HTML}{000000}

\mathchardef\mhyphen="2D

\newcommand{\roughly}{\ensuremath{ {\sim}\,} }

\newlength{\dhatheight}

\newcommand{\code}[1]{\texttt{#1}\xspace}

\newcommand{\secref}[1]{Section~\ref{sec:#1}}

\newcommand{\tabref}[1]{Table~\ref{tab:#1}}

\newcommand{\figref}[1]{Figure~\ref{fig:#1}}

\newcommand{\bandvar}[2][]{%
  \ifthenelse{\isempty{#1}}{\var{#2}}{\var{#2\_#1}}%
}

\newcommand{\var}[1]{\ensuremath{\texttt{\MakeUppercase{#1}}}\xspace}

\providecommand\physrep{\ref@jnl{Phys.~Rep.}}%
\providecommand\apjs{\ref@jnl{ApJS}}%
\providecommand{\jcap}{\ref@jnl{JCAP}}%

\shorttitle{Multi-Wavelength (VIJHK) TRGB Calibration from Globular Clusters}
\shortauthors{Cerny et al.}

\begin{document}

\title{Multi-Wavelength, Optical (VI) and Near-Infrared (JHK) Calibration\\ of the \\Tip of the Red Giant Branch Method \\based on \\Milky Way Globular Clusters}

\author[0000-0003-1697-7062]{William Cerny}
\affiliation{Department of Astronomy and Astrophysics, University of Chicago, \\ 5640 S. Ellis Ave., Chicago IL 60637, USA}
\affiliation{Kavli Institute for Cosmological Physics, University of Chicago, \\ 5640 S. Ellis Ave., Chicago, IL 60637, USA}
\author[0000-0003-3431-9135]{Wendy L. Freedman}
\affiliation{Department of Astronomy and Astrophysics, University of Chicago, \\ 5640 S. Ellis Ave.,  Chicago IL 60637, USA}
\affiliation{Kavli Institute for Cosmological Physics, University of Chicago, \\ 5640 S. Ellis Ave., Chicago, IL 60637, USA}
\author[0000-0002-1576-1676]{Barry F. Madore}\textbf{}
\affiliation{The Observatories, Carnegie Institution for Science\\
 813 Santa Barbara St., 
Pasadena CA 91101, USA}
\author{Finian Ashmead}
\affiliation{Department of Astronomy and Astrophysics, University of Chicago, \\ 5640 S. Ellis Ave., Chicago IL 60637, USA}
\author[0000-0001-9664-0560]{Taylor Hoyt} 
\affiliation{Department of Astronomy and Astrophysics, University of Chicago, \\ 5640 S. Ellis Ave.,
Chicago IL 60637, USA}
\author[0000-0002-0119-1115]{Elias Oakes} 
\affiliation{Department of Astronomy and Astrophysics, University of Chicago, \\ 5640 S. Ellis Ave., Chicago IL 60637, USA}
\author[0000-0001-6532-6755]{Nhat Quang Hoang Tran}
\affiliation{Department of Astronomy, University of Texas at Austin, 2515 Speedway, Stop C1400\\
Austin, Texas 78712-1205 , USA}
\author{Blake Moss}
\affiliation{Department of Astronomy and Astrophysics, University of Chicago, \\ 5640 S. Ellis Ave.,  Chicago IL 60637, USA}

\correspondingauthor{William Cerny}
\email{williamcerny@uchicago.edu}

\begin{abstract}
Using high precision ground-based photometry for 46 low-reddening Galactic globular clusters, in conjunction with Gaia DR2 proper motions for member star selection, we have calibrated the zero point of the tip of the red giant branch (TRGB) method at two optical ($VI$) and three near-infrared ($JHK$) wavelengths. In doing so, we utilized the sharply-defined zero-age horizontal branch (ZAHB) of these clusters to relatively calibrate our cluster sample into a composite color-magnitude diagram spanning a wide range of metallicities, before setting the absolute zero point of this composite using the geometric detached eclipsing binary distance to the cluster $\omega$~Centauri. The $I-$band zero point we measure
[$M_I = -4.056 \pm 0.02 \text{ (stat})  \pm 0.10 \text{ (sys)} $]
agrees to within one sigma of the two previously published independent calibrations, using TRGB stars in the LMC [$M_I = $ -4.047~mag; Freedman et al. 2019, 2020] and in the maser galaxy NGC~4258 [$M_{F814W} = $ -4.051~mag; Jang et al. 2020]. We also find close agreement for our $J,H,K$ zero points to several literature studies.

\end{abstract}



\section{Introduction}
\label{sec:intro}

For over a century, the brightest stars in globular clusters have played a pivotal role in our understanding of the scale size of our universe. These include the ``Great Debate" \citep{shapley_curtis_1921} over the size (and the uniqueness) of our universe where Harlow Shapley used the brightest giants for determining distances to globular clusters.\footnote{Shapley was on the wrong side of history in that application; he assumed that there were no corrections for dust, and that his Cepheid zero-point calibration was correct (But, of course, there is plenty of dust in the plane of our Milky Way; and only later was it found that there are two types of Cepheids, differing by a factor of 2 in luminosity).} Several decades later, by using newly-developed red-sensitive emulsions on photographic plates, Walter Baade \citep{baade_1944} was able to resolve tip of the Red Giant Branch (TRGB) stars in the Local Group galaxies  M31, M32 and  NGC~205, resulting in a major recalibration of the distance scale. Measuring the magnitude difference between tip stars in Galactic globular clusters and RR Lyrae variables in the same clusters led Baade to revise  Hubble's measurement of the expansion rate by a factor of two. 
The TRGB method also played an important role in ending the ``factor-of-two" debate over the Hubble constant that remained unresolved during the last three decades of the 20th century: the HST Key Project on the extragalactic distance scale used TRGB distances to galaxies both as a consistency check on the zero point of the Cepheid calibration, and to undertake a differential test for metallicity effects on the Leavitt Law \citep[e.g., see][and references therein]{freedman_2001}.
\par The value of the Hubble constant has recently once again become a subject of debate \citep[e.g.][]{freedman_2017, efstathiou_2020}. Planck measurements of the cosmic microwave background (CMB) temperature and polarization power spectra 
give a value of $H_o = $ 67.4 $\pm$ 0.5 km/s/Mpc, under the assumption of $\Lambda$CDM \citep{planck_2018}. On the other hand, Cepheids used to calibrate Type~Ia supernovae give values of the Hubble constant around 74 km/s/Mpc \citep{riess_2019}. Taking the quoted uncertainties at face value, the two numbers disagree at greater than 4-sigma. If this result holds the test of time, it may be indicating new physics beyond the standard cold dark matter ($\Lambda$CDM) model. However, recent {\it HST} measurements of the TRGB \citep{freedman_2019, freedman_2020} provide a calibration of the Type Ia supernovae distance scale, and find a value of H$_o$ = 69.6 $\pm$0.8 (stat) $\pm$ 1.7 (sys) km/sec/Mpc. This value is consistent with the Planck measurements at 1-sigma, but still consistent with the Cepheid distance scale to better than 2-sigma. The zero-point calibration in this case was based on a detached eclipsing binary measurement to the Large Magellanic Cloud (LMC) \citep{pietrzynski_2019}. 
\par In this paper, we undertake an entirely independent calibration of the TRGB method. This calibration is based on optical ($VI$) and near-infrared ($JHK$) photometry of red giant stars in a sample of 46 globular clusters in the Milky Way. These clusters have a range of metallicities that span those observed in the halos of nearby galaxies for which TRGB distances can be measured. The organization of this paper is as follows. In \secref{previous}, we review previous calibrations of the TRGB using Galactic globular clusters. In \secref{data}, we describe the cluster data used for this work, and our procedure for identifying cluster member stars algorithmically using proper motion data from \Gaia DR2. In \secref{calibration}, we utilize the ZAHB to derive {\it relatively} calibrated cluster distances for all clusters in our sample, before finally establishing an absolute zero-point for these distances using the DEB distance to $\omega~Cen$.

\section{Previous Calibrations of the TRGB Method}
\label{sec:previous}

The development of the TRGB as an accurate distance indicator improved significantly with the availability of CCD detectors.  Obtaining $VI$ CCD data for six Milky Way globular clusters, \citet{da_costa_armandroff_1990}  used distances based on theoretical horizontal branch models to calibrate the luminosities of RR Lyrae stars. The first application of this Galactic calibration to the extragalactic distance scale was undertaken by \citet{lee_freedman_madore_1993}. These authors compared the TRGB distances to 10 nearby galaxies with those based on RR Lyrae stars and Cepheids, and quantitatively demonstrated the power of the TRGB method for measuring distances to nearby galaxies, a method comparable in accuracy to the Cepheid Leavitt law.
\par \citet{ferraro_1999} updated the Milky Way TRGB calibration, based on a larger sample of 60 globular clusters. These authors adopted the zero-age horizontal branch (ZAHB) as a standard candle upon which to base their distances, avoiding altogether the use of RR Lyrae variables. This method takes advantage of the simplicity of the measurement of the horizontal branch, as compared to measurement of the period-luminosity-metallicity relation for variable RR Lyrae stars. As these authors noted, the mean RR~Lyrae locus is not coincident with the (fainter) position of the ZAHB. More recently, the development of a technique to measure accurate distances using detached eclipsing binaries (DEBs) provided a new and independent means of calibrating the TRGB in Galactic globular clusters. \citet{bellazzini_2001, bellazzini_2004} based their calibration of $\omega~Cen$ and $47~Tuc$ on the detached eclipsing binary distance to $47~Tuc$ \citep{thompson_2001}. In a later study, \citet{rizzi_2007}  used the horizontal branches of five Local Group galaxies (IC~1613, NGC~185, Fornax, Sculptor and M33) for their calibration of the TRGB. 

\par These calibrations of the TRGB have generally been in good agreement, leading to an absolute magnitude for the I-band TRGB in the range of -4.00 to -4.05~mag. We next describe our methodology for measuring the TRGB based on a homogeneous sample of Milky Way globular clusters.

\begin{deluxetable*}{l c c c }
\tablecolumns{4}
\tabletypesize{\small}
\tablecaption{\label{tab:properties}
Adopted Metallicities, Derived Distance Moduli \& Reddenings for the 46 Selected Galactic Globular Clusters}
\tablehead{
\colhead{Cluster} &  \colhead{[Fe/H]$_{\rm H10}$} & \colhead{$(m-M)_0$} & \colhead{$E(B-V)$}}
\startdata
NGC 6362 & -0.99 & 14.47 & 0.08 \\
NGC 6723 & -1.10 & 14.69 & 0.06 \\
NGC 2808 & -1.14 & 15.10 & 0.17 \\
NGC 1851 & -1.18 & 15.42 & 0.03 \\
NGC 362 & -1.26 & 14.76 & 0.04 \\
NGC 1261 & -1.27 & 16.10 & 0.01 \\
NGC 6864 (M75) & -1.29 & 16.67 & 0.14 \\
\textbf{NGC 5904} (M5) & \textbf{-1.29} & \textbf{14.41} & \textbf{0.03} \\
NGC 288 & -1.32 & 14.83 & 0.01 \\
NGC 6218 (M12) & -1.37 & 13.56 & 0.17 \\
\hline
NGC 6981 (M72) & -1.42 & 16.11 & 0.05 \\
NGC 6934 & -1.47 & 16.01 & 0.10 \\
NGC 6229 & -1.47 & 17.43 & 0.02 \\
NGC 6584 & -1.50 & 15.69 & 0.10 \\
NGC 5272 (M3) & -1.50 & 15.08  & 0.01 \\
NGC 7006 & -1.52 & 18.05 & 0.07 \\
\textbf{NGC 5139} ($\omega \rm Cen$) & \textbf{-1.53} & \textbf{13.678} & \textbf{0.12} \\
NGC 6205 (M13) & -1.53 & 14.30 & 0.02 \\
IC4499 & -1.53 & 16.52 & 0.20 \\
NGC 6752 & -1.54 & 13.04 & 0.05 \\
NGC 5986 & -1.59 & 15.15 & 0.28 \\
NGC 3201 & -1.59 & 13.49 & 0.23 \\
NGC 1904 (M79) & -1.60 & 15.57 & 0.02 \\
NGC 7089 (M2) & -1.65 & 15.41 & 0.04 \\
NGC 5286 & -1.69 & 15.36 & 0.23 \\
\hline
NGC 6093 (M80) & -1.75 & 15.01 & 0.18 \\
NGC 7492 & -1.78 & 16.77 & 0.03 \\
NGC 4147 & -1.80 & 16.37 & 0.02 \\
\textbf{NGC 6541} & \textbf{-1.81} & \textbf{14.40} & \textbf{0.12} \\
NGC 5634 & -1.88 & 17.02 & 0.05 \\
NGC 5897 & -1.90 & 15.43 & 0.11 \\
NGC 5824 & -1.91 & 17.52 & 0.14 \\
NGC 6809 (M55) & -1.94 & 13.53 & 0.12 \\
NGC 5466 & -1.98 & 15.98 & 0.01 \\
NGC 6779 (M56) & -1.98 & 14.99 & 0.22 \\
NGC 5694 & -1.98 & 17.75 & 0.09 \\
NGC 6101 & -1.98 & 15.78 & 0.10 \\
\hline
NGC 6397 & -2.02 & 11.90 & 0.17 \\
NGC 5024 (M53) & -2.10 & 16.30 & 0.02 \\
NGC 2419 & -2.15 & 19.61 & 0.08 \\
Terzan8 & -2.16 & 17.10 & 0.12 \\
NGC 4590 (M68) & -2.23 & 15.02 & 0.05 \\
NGC 7099 (M30) & -2.27 & 14.52 & 0.04 \\
NGC 5053 & -2.27 & 16.19 & 0.01 \\
\textbf{NGC 6341 (M92)} & \textbf{-2.31} & \textbf{14.59} & \textbf{0.02} \\
NGC 7078 (M15) & -2.37 & 15.11 & 0.08 \\
\enddata
\vspace{-3em}
\tablecomments{[Fe/H] values are from H10. }
\end{deluxetable*}

\section{Cluster Data}
\label{sec:data}

\subsection{Cluster Selection and Optical Data}
We initially selected our sample of globular clusters from the catalogs presented in \citet{S19}, hereafter ``S19," which included 48 clusters in total. These clusters were originally chosen in S19 based on the availability of high-quality observations in all five of the optical UBVRI filter bands, where the criterion for quality was based on the ability to derive a precise and accurate absolute calibration for the photometry. As S19 describes, homogeneous reduction of the observations was performed over more than 80,000 individual CCD images using the DAOPHOT, ALLFRAME, and ALLSTAR suites of programs \citep{Stetson_1987, allstar_1992, allframe}.  However, since we only sought to calibrate the TRGB in the V and I bands, we opted to eliminate 11 high-reddening ($E(B-V) > 0.25$) clusters from this initial S19 sample, replacing them with nine other bright, low-reddening NGC globular clusters available from the CADC archive\footnote{www.cadc-ccda.hia-iha.nrc-cnrc.gc.ca/en/community/STETSON} as of September 2020 -- all processed in the same homogeneous manner -- to build our final sample of 46 clusters. These 9 clusters were not in S19 because data in all of the UBVRI bands were not available or not yet reduced at the time of publication. Hereafter, we include these clusters when referring to S19 throughout this work.  We list all clusters included in this study in \tabref{properties}.
\par As S19 emphasizes, the resulting photometry for these Galactic globular clusters is well-suited to be utilized in conjunction with the high-precision photometry and proper motion measurements provided by \Gaia DR2. To combine these catalogs, we performed a cross-match between the S19 positional data and the complete \Gaia DR2 database using a 0.5~arcsec matching radius employing the CDS {\code {xMatch}} service.\footnote{http://cdsxmatch.u-strasbg.fr/} We then retained only the matched stars that also had proper motion measurements. We did not use any parallax information in this study, anticipating superior determinations in the upcoming \Gaia Early Data Release 3 (EDR3) release. We then further applied conservative quality cuts based on the DAOPHOT parameters $\chi$ and $\rm sharp$. In particular, we applied $\chi < 5$ and $|\rm sharp| < 1$, and after selecting member stars for each cluster (described in the following subsection), we additionally removed all stars within $min$(core radius, 1~arcmin) of angular separation from the cluster centroids, utilizing core radii and centroids from the \citet{H10} \space catalog of globular cluster properties. Hereafter, we refer to this catalog, most recently updated in 2010, as H10. These cuts are intentionally generous so as to be as complete as possible, while also maintaining a reasonably high level of purity.

\begin{figure*}
\center

\includegraphics[width=\textwidth]{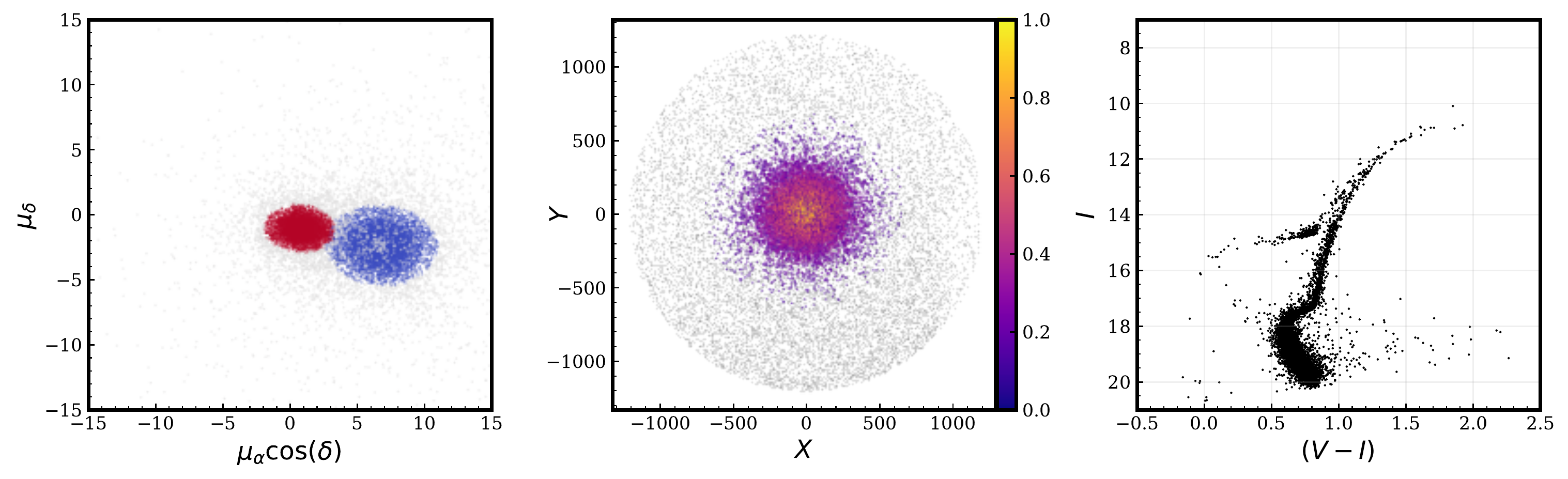}
\caption{
    Diagnostic plots generated when applying our Gaussian Mixture Model membership to the cluster NGC~362 (a cluster with two distinct signal components).
   (Left) 2D distribution of the \Gaia DR2 proper motion measurements for all stars cross-matched between the catalogs provided by S19 and \Gaia DR2 within 1.25 tidal radii of the cluster centroid (H10;\citealt{V19}). The blue ellipsoid corresponds to stars identified as likely cluster members, while the red ellipsoid corresponds to stars that are likely members of the SMC. Points drawn in grey are foreground or background stars that the model predicts do not belong to either of the two ``signals" in the S19 photometric catalogs. Grey stars that appear in the region dominated by the blue ellipsoid are individual stars that featured proper motion measurements that are approximately consistent with the cluster's systemic mean proper motion, but were excluded from being identified as likely cluster member stars due to large spatial separation from the cluster centroid. (Center) Spatial distribution of NGC~362 in planar $(X,Y)$ coordinates. Stars displayed in color are those identified as member stars by the Gaussian Mixture Model, while stars in grey correspond to those identified as either SMC or foreground/background stars. In general, stars are increasingly likely to be deemed members by the model as their positions approach the cluster centroid; however, we note that the precision and accuracy of the \Gaia proper motions decreases significantly in the crowded cores of the majority of the clusters studied in this work.
   (Right) Color-magnitude diagram (CMD) of the identified member stars for NGC~362, with no distance, reddening, or extinction corrections applied. In order to mitigate crowding effects, we exclude stars within $min$(core radius, 1 arcminute) of the cluster centroid when constructing and analyzing CMDs in this work. 
}
\label{fig:GMM}
\end{figure*}

\subsection{Selecting Cluster Member Stars}

\begin{figure*}
\center
\includegraphics[width=\textwidth]{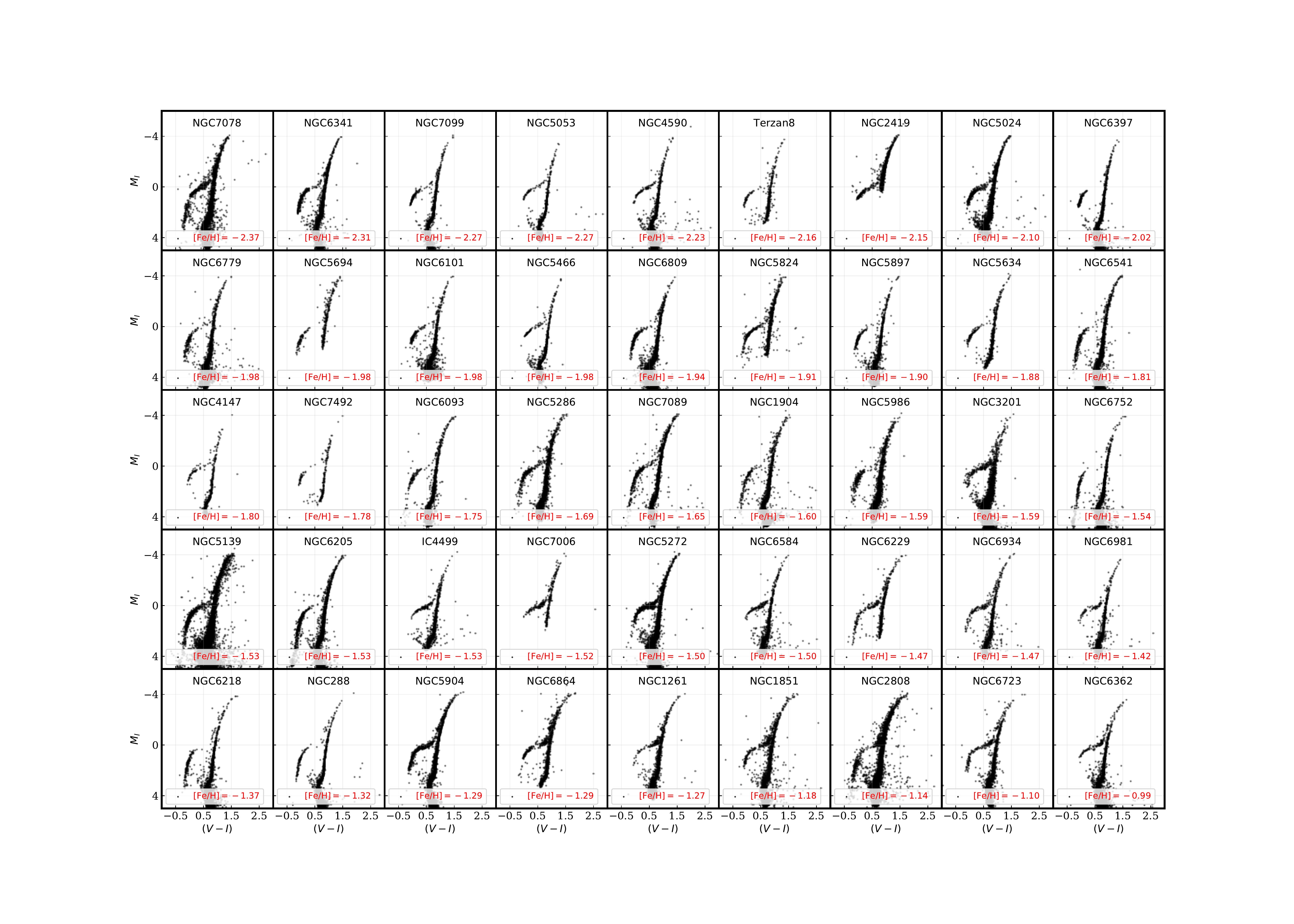}
\caption{46 Galactic Globular Cluster Optical CMDs used in this study, individually displayed and ordered by increasing metallicity (shown in red) from the top-left to bottom-right. Adopted distance moduli and reddenings are given in Table 1. We exclude NGC~362 in this plot, as an identical CMD is displayed in \figref{GMM}. Clusters with ``missing" main sequences correspond to some of the more distant globular clusters with stellar populations that extend to magnitudes fainter than the \Gaia limiting magnitude of $G\roughly21$~mag (e.g., NGC~2419, NGC~5694 and NGC~7006).}

\label{fig:all_optical_cmds}
\end{figure*}

\label{sec:membership}
In order to isolate cluster member stars from foreground and background stars, we utilized the {\code{scikit-learn}} \citep{sklearn} implementation of the Gaussian Mixture Model clustering algorithm, inspired by the methods presented in \citet{Fierro:19} and \citet{V19}.
Briefly, a Gaussian Mixture Model is a fast and robust supervised clustering algorithm capable of fitting a multi-dimensional Gaussian to a fixed number of classification components; applying the algorithm allows for the probabilistic labelling of all sources into these distinct components.
\par The clustering algorithm we applied takes in four dimensions as inputs: the two spatial coordinates $(X, Y)$ derived for each individual cluster in S19 and two proper motion vector components ($\mu_{\alpha}cos(\delta),\mu_{\delta})$ taken from the \Gaia catalog. The $(X,Y)$ coordinates are chosen for the same reason presented in \citet{Fierro:19}, namely that they mitigate the potential for the mixture model to treat clusters at varying declinations differently due to spherical geometry; here, instead, the planar $(X,Y)$ coordinates were derived in S19 so as to fix a consistent radial scale centered at the position of each cluster. For a given cluster, we begin by pre-processing these measurements using {\code{scikit-learn}}'s {\it RobustScaler} method over these four input features, which scales the feature values to equal orders of magnitude based on their inter-quartile ranges in order to weight them equivalently in membership determinations. We do not utilize \Gaia parallaxes or parallax-derived quantities in these models, as prior attempts in the literature to use them for cluster member star determinations resulted in reduced machine learning model efficiency (eg, \citealt{Fierro:19}), and since these measurements can be imprecise in densely populated regions of clusters. We also note that \Gaia radial velocities remain unavailable for the vast majority of stars in most clusters, and thus we do not utilize these measurements for our mixture model classification. 
\par In using the Gaussian Mixture Model algorithm, we generally impose that there are two mixture model components for each cluster's dataset: one representing signal based on the cluster's spatial distribution and proper motion vector, and the second ``background.'' After the algorithm is run over each cluster, we visually inspect the resulting spatial (coordinate) distribution, proper motion vector-point diagrams, and color-magnitude diagrams. For a limited number of clusters where visual inspection elucidated high levels of  contamination from foreground/background stars, we adjusted the number of expected components and re-run the algorithm in order to provide a better membership classification for all stars in the data. Such a change proved to be particularly important for cluster fields with other nearby astronomical objects; for example, in the case of the cluster NGC~362, the close proximity of the Small Magellanic Cloud (SMC) in projected (2D) coordinate separation and its subsequent contribution to the other two clusters' catalogs demanded that we increase the number of expected mixture model components to three, reflecting that of the SMC, the desired cluster, and foreground/background stars. One key advantage of using the Gaussian mixture model, however, is that we need not specify the expected centroid or mean proper motion signal of each component -- these are automatically derived, and our model only assumes that each component is well-represented by a multi-dimensional Gaussian over the input features. Additionally, in a limited number of additional cases where the proper motion signal for the cluster is less prominent, we apply exceedingly conservative cuts on the proper motion components as priors; these priors are chosen so as to guide the algorithm's membership classifications without excluding any potential member stars. We emphasize that by not requiring parallaxes in our membership procedure, our catalogs 
are generally more complete than those provided by \citet{kinematics}, featuring tens to thousands of additional member stars (typically at fainter magnitudes) without compromising the purity of the color-magnitude diagram for each cluster. 

\par In \figref{GMM}, we visually depict diagnostic results of the cluster membership process described above for the cluster NGC~362; this cluster is chosen to demonstrate a case when a three-component mixture model is used to isolate the cluster signal from that of the SMC and the foreground/background field of stars. In the left-hand panel, we plot the distribution of proper motion measurements ($\mu_{\alpha}cos(\delta),\mu_{\delta})$  for all stars cross-matched between S19 and \Gaia DR2 within 1.25 tidal radii of the cluster centroid and that pass our initial quality cuts. Points in blue are those sources identified as likely cluster members, whereas points in red are likely SMC stars; light grey points correspond to foreground or background stars, which belong to neither system. In the center panel, we plot the spatial distribution of likely member stars in planar $(X,Y)$ coordinates, colored by membership probability (normalized to unity).  Field stars identified as non-members are plotted as grey points. Lastly, in the right-hand panel, we plot the $I$ vs. $(V-I)$ color--magnitude diagram for the likely member stars for the cluster (uncorrected for reddening or extinction). The quality of the S19 photometry and the purity of the member-star sample is evident, with minimal contamination visible in the upper red giant branch and asymptotic giant branch. In  \figref{all_optical_cmds}, we present comparable $I$ vs. $(V-I)$ color--magnitude diagrams for all clusters analyzed in this study.

\subsection{Near-Infrared (JHK) Data}
In order to extend our TRGB measurement to the near-infrared, we utilize the high-precision $J,H,K$ photometry from the 2 Micron All-Sky Survey (2MASS; \citealt{2mass_2006}). To construct our stellar catalog, we perform a cross match between the full \Gaia DR2 catalog and the 2MASS Point-Source Catalog using the CDS {\it xMatch} service, with a 0.5'' matching radius. We then sub-select all sources within this cross-match that were also identified in the optical catalog (see previous subsection), by selecting stars based on their \Gaia DR2 source IDs.

\begin{figure*}
\center

\includegraphics[width=\textwidth]{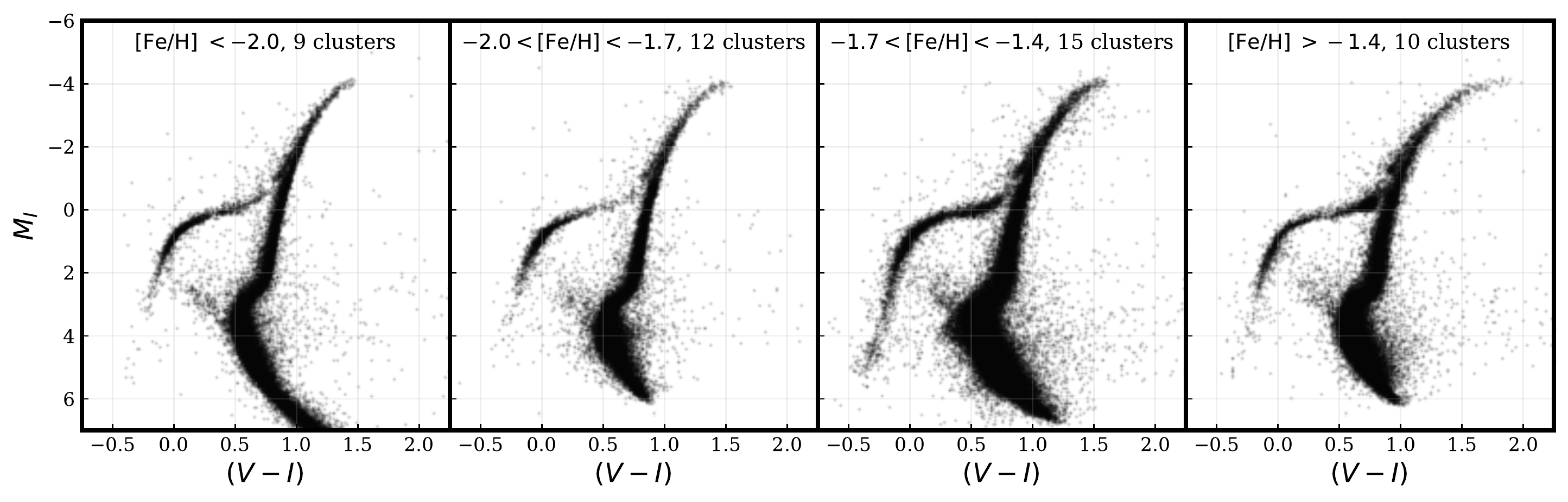}
\caption{
    Optical-wavelength ($I$ versus $V-I$) composite color-magnitude diagrams for the four metallicity bins used during the relative alignment process in \secref{calibration}, corrected for distance and reddening/extinction, and brought onto the absolute zero-point set by the DEB distance to $\omega~Cen$. The clusters corresponding to each bin are listed in \tabref{properties}. These composite CMDs reveal the precise alignment of the cluster ZAHBs achieved by the process described in \secref{calibration}. While some contamination is visible at fainter magnitudes, the four composites individually exhibit the rapid drop-off in density expected expected from the TRGB at  $I_{0} \lesssim - 4.0$~mag. Such a feature would not have been nearly as distinct from any individual cluster excluding $\omega~Cen$.
    }

\label{fig:metal_bins}
\end{figure*}

\section{Calibration of Cluster Distances}
\label{sec:calibration}

\subsection{Initial Reddening and Distance Corrections}
In order to bring the individual 46 clusters onto a consistent relative calibration, we first begin by converting all measurements to absolute magnitudes. 
To do so, we initially adopted the apparent visual distance moduli and color excesses $E(B-V)$ from the H10 catalog. We then converted the former to true distance moduli assuming a \citet{C89} reddening law with $R_{V}$ = 3.1 for all clusters. Next, we applied an extinction correction in each band by adopting the H10 reddening values and the following total-to-selective absorption ratios for the Johnson-Cousins $BVI$ filters: $ R_B = 4.145, R_V = 3.1, R_I = 1.485$  from \citet{C89}. We later adopt $ R_J = 0.874, R_H = 0.589, R_K = 0.353$ \citep{C89} when correcting the photometry from 2MASS; however, reddening corrections in the NIR were not applied until after we derived our final true distance moduli and $E(B-V)$ color excess values using the optical data.

\subsection{Securing a Relative Cluster Calibration}
With these corrections in place, we next sought to bring all the clusters onto a consistent relative calibration. To do so, we began by splitting the cluster sample into four distinct metallicity bins based on the metallicity values reported in H10.\footnote{https://www.physics.mcmaster.ca/~harris/mwgc.dat} These four bins correspond to the metallicity ranges: $[\rm Fe/H] < -2.0$, $ -2.0 \leq [\rm Fe/H]  < -1.7$, $ -1.7 \leq [\rm Fe/H]  < -1.4$, and $[\rm Fe/H] > -1.4$~dex, containing nine, twelve, fifteen, and ten clusters per bin, respectively (see \tabref{properties}, where horizontal divisions divide the clusters by bin). Within each bin, we selected a single bright, nearby, low-extinction cluster with a sharply-defined and  well-populated blue horizontal branch (based on visual inspection of all CMDs) as a ``reference" cluster which we utilized to (relatively) calibrate the remaining clusters within that bin. We indicate the reference cluster for each bin in bold in \tabref{properties}. The practical advantage of binning the clusters in this manner is to group clusters with similar horizontal branch morphologies, allowing for algorithmic, likelihood-based alignment of their HBs.
\par In order to align the clusters within each bin to the reference cluster, we first constructed a fiducial horizontal branch by fitting a finely-interpolated line along the sharply-defined lower envelope of the blue horizontal branch in the $V,I$ bands.
With the derived blue horizontal branch,  we then applied a simple maximum-likelihood grid search technique in order to bring the lower envelopes of the remaining clusters into the same relative calibration (based on the $V,I$ bands). 
Within a given bin,  we initially assumed the calculated true distance modulus and reddening values from H10, and stepped through potential distance modulus offsets in a range [-0.25 to 0.25~mag] in intervals of 0.01~mag in order to find the distance modulus offset which best brought the horizontal branch of each cluster into alignment with the fiducial HB. 
The likelihood function we sought to maximize for this grid search was defined as the raw number of horizontal branch stars for each cluster that fell between the fiducial lower envelope from the reference cluster and a parallel branch displaced .05~mag brighter, factoring in photometric uncertainties for each star.  
For each value of the distance modulus shift in the interval, we added the distance modulus offset to all stars' $I-$band magnitudes, and re-calculated the number of stars lying in the horizontal branch region defined above. In some cases where the morphology of a given cluster varied to a non-negligible degree with respect to the reference cluster for its bin (for example, when the horizontal branch naturally truncates at redder photometric colors), we restricted the color range of the fiducial to match that of the given cluster when applying the likelihood procedure; this minimizes the possibility of contaminant stars affecting the likelihood determination.  Additionally, in a number of clusters, we found evidence for clear misalignment of the lower red giant branch and subgiant branch of clusters compared to the (low reddening) reference cluster far beyond what might be expected due to differences in metallicity alone within the small ($\roughly 0.3$~dex) metallicity ranges that define each bin. Therefore, for these cases, we iteratively adjusted the $E(B-V)$ color excess of each cluster (recalculating the extinction in both bands and the reddening) in intervals of 0.01~mag and took note of the pair of (reddening, distance modulus offset) which resulted in both the clearest alignment of the blue horizontal branch and the maximization of the number of stars lying within the bounds constructed using the fiducial horizontal branch. 

\par In more metal-rich clusters where a reasonably-populated horizontal branch was not present, we instead fit to the lower edge of the red clump; in our sample, the metallicity range in our most metal-rich bin is narrow enough that significant variation (metallicity dependent effects) in the red clump is not expected across clusters. In particular, we utilized NGC~1261, which is low reddening ($E(B-V) = 0.01$~mag) and features both a well populated red horizontal branch/clump and enough of a blue horizontal branch to construct a fiducial ZAHB lower envelope. As a result of doing so, this cluster could be used as reference for clusters both with and without blue horizontal branches. 
\par We have excluded the bright, well-populated, cluster 47 Tucanae from our analysis, as it is significantly more metal-rich  ($\rm [Fe/H] = -0.72$; \citealt{H10}) than the other clusters considered in this study, which are bounded by a metallicity of $\rm [Fe/H] = -0.99$ (NGC~6362). This metallicity gap renders it nearly impossible to bring 47 Tuc onto a relative calibration with all the other clusters, as metallicity effects in the red clump appear to become significant. Additionally, while 47 Tuc also has a well-constrained geometric distance from detached eclipsing binaries \citep{thompson_2020}, we did not use this measurement to set the absolute zero-point in this work due to both the aforementioned metallicity gap and the lack of a horizontal branch for this cluster. However, these issues will be resolved with accurate parallax-based distances to these clusters with the forthcoming \Gaia Data Release 3, allowing for all clusters to be brought onto a self-consistent geometric distance scale (see \secref{discussion}). 
\par The final result of these procedures is a composite color-magnitude diagram for each of the four metallicity bins, shown in \figref{metal_bins}, with the clusters in each bin brought onto a uniform relative calibration. We estimate the (statistical) uncertainty of cluster distance moduli derived from this intra-bin cluster horizontal branch alignment process as $\pm 0.03$~mag, corresponding to the typical full width of the maximum-likelihood response peak about the most probable value for a cluster's vertical shift.
\subsection{Absolute Calibration}

We set the absolute zero-point for the relatively-calibrated composite of clusters constructed above to a DEB-based geometric distance to the cluster $\omega~Cen$. \citet{thompson_2001} measured an apparent distance modulus to the system of $(m-M)_{V} = 14.05 \pm 0.11$~mag. In the previous subsection, we found a best-fit color excess value for $\omega~Cen$ of $E(B-V) = 0.12$~mag\footnote{For reference, \citet{thompson_2001} adopted a value $E(B-V) = 0.13$~mag, as derived from \citet{schlegel_1998} maps.} and thus we recalculate the true distance modulus to the system to be
$\mu_{0} = 14.05 - (3.1 * 0.12) = 13.678$~mag. Adopting this DEB-based distance to $\omega~Cen$, we re-zero the third metallicity bin containing $\omega~Cen$. Then, we simply applied our maximum-likelihood HB-fitting technique three times more in total, once to each metallicity bin, in order to bring each of the remaining three bins' composite HBs onto the newly-set zero point provided by the $\omega~Cen$ DEB distance. In doing so, we found that only small (0.01 - 0.04~mag) shifts were required. We estimate the error associated with this inter-bin alignment to be $\pm 0.02$~mag, again based on the width of the maximum-likelihood response.
\par We emphasize that our procedure for multi-cluster alignment is fully independent of RGB stars, instead only relying on the zero-age horizontal branch feature located 4~mag below the TRGB feature. 

\par We report our derived values of true distance modulus and color excess for all 46 clusters (calculated by first assuming the H10 value, and  then applying our derived distance modulus and/or reddening shifts, if applicable) in \tabref{properties}. 
For each true distance modulus, $(m-M)_{0}$, we attribute a statistical error based on three sources of uncertainty: (1) a 0.015~mag statistical uncertainty associated with the alignment in color of a given cluster against $\omega~Cen$ due to an empirical 0.01~mag uncertainty in the $E(B-V)$ color excess, propagated to the the $I-$band; (2) a 0.03~mag statistical uncertainty arising from the intra-bin alignment described in the previous subsection; and (3), a 0.02~mag statistical uncertainty arising from the inter-bin alignment described above in the current subsection. Adding the respective components of the statistical errors in quadrature, we assign an uncertainty to our reported true distance moduli of $ (m-M)_{0}  \pm 0.039 \text{ (stat})  $~mag. We return to the issue of systematic errors in Section \ref{sec:measurement} below.

\begin{figure*}

\center
\includegraphics[width=.34\textwidth]{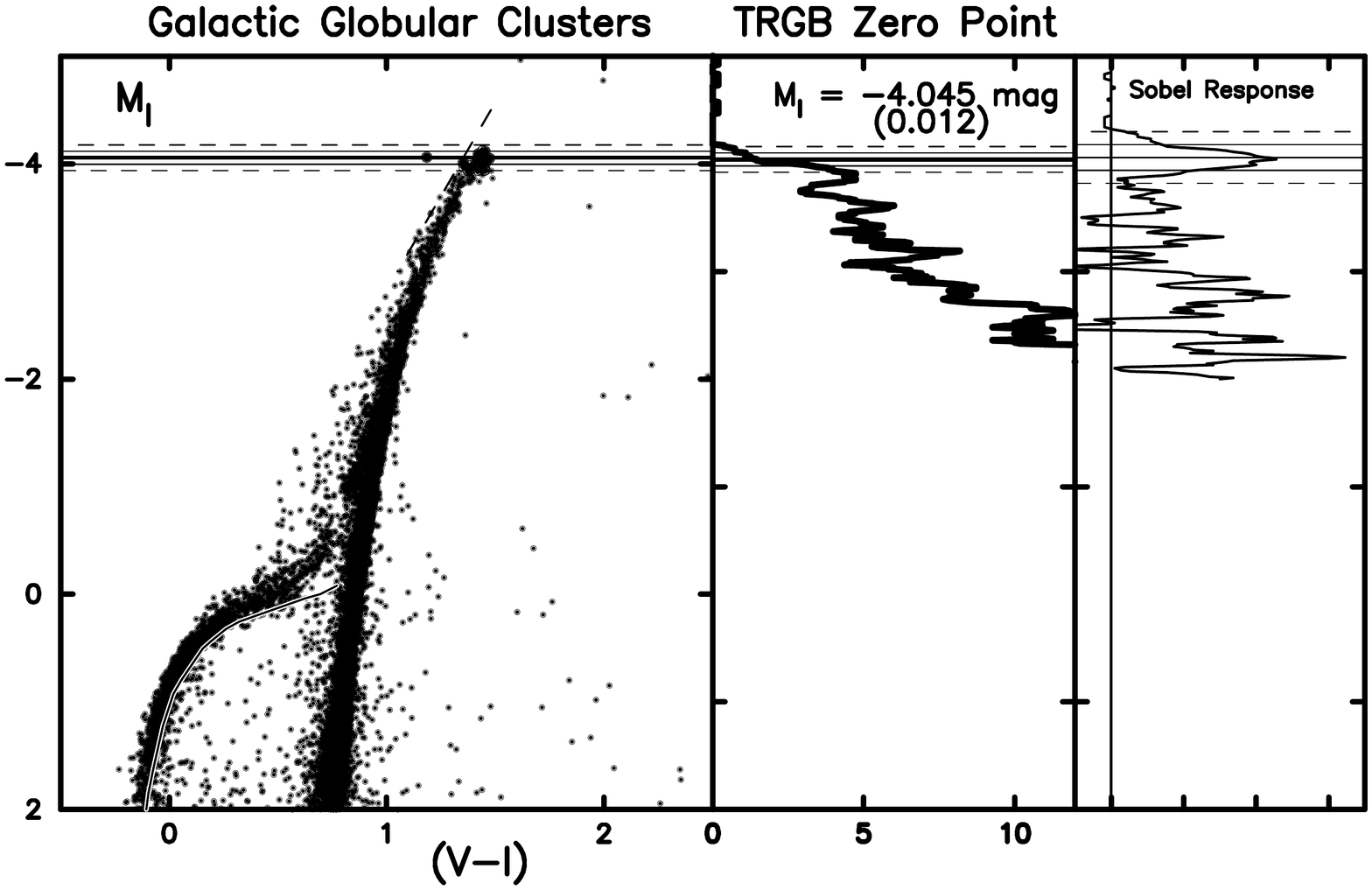}
\includegraphics[width=.34\textwidth]{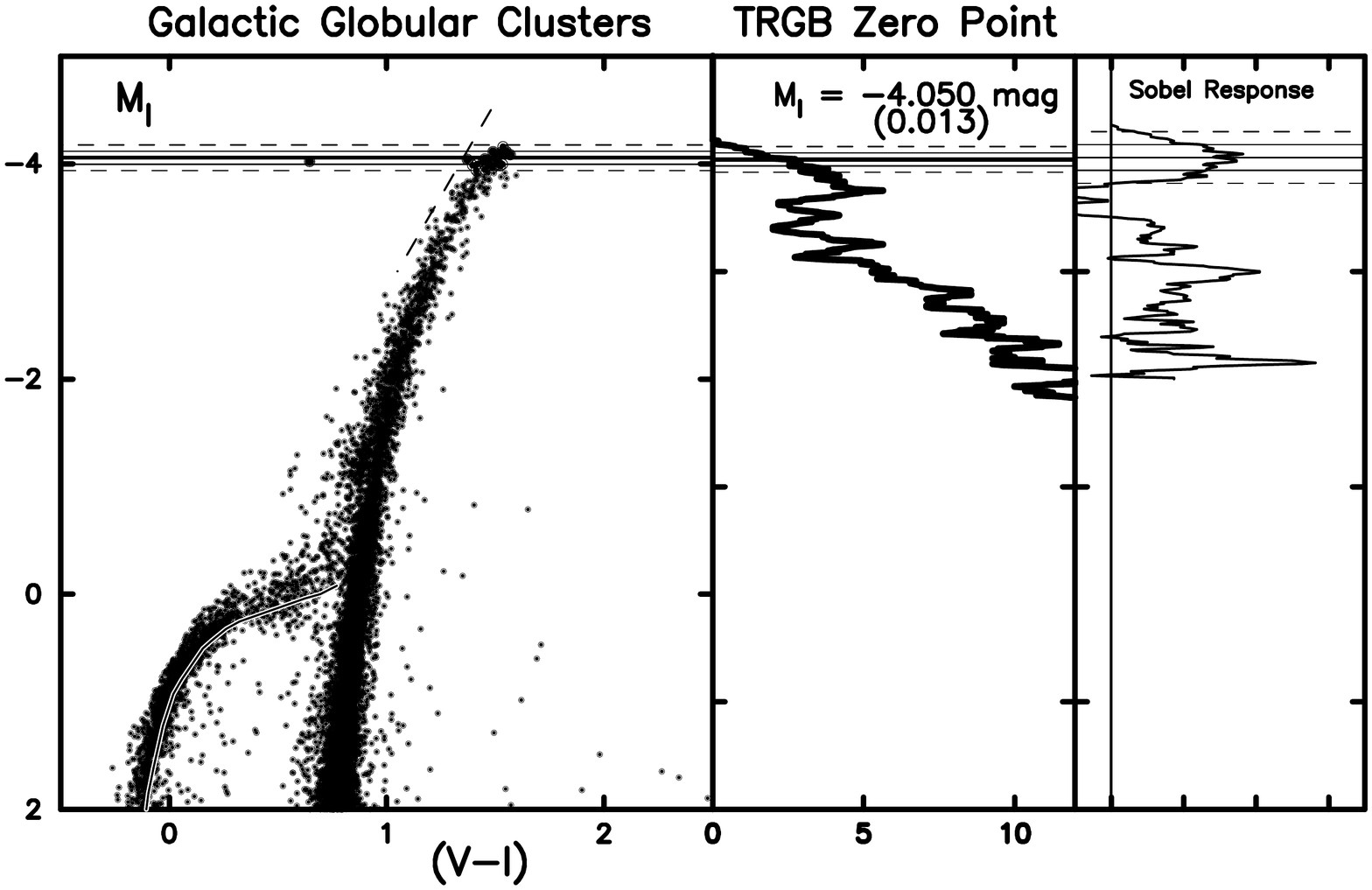}
\includegraphics[width=.34\textwidth]{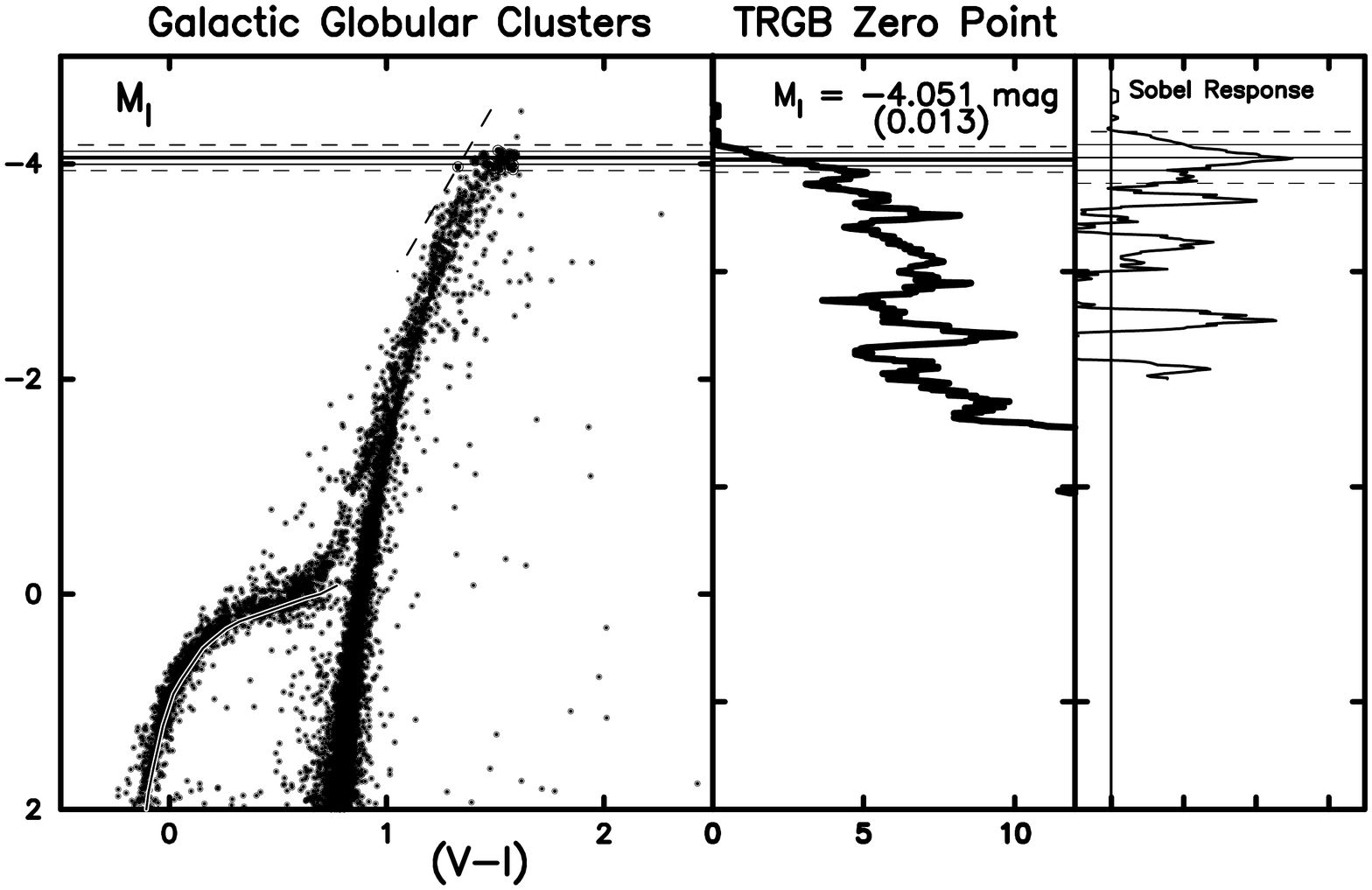}
\includegraphics[width=.34\textwidth]{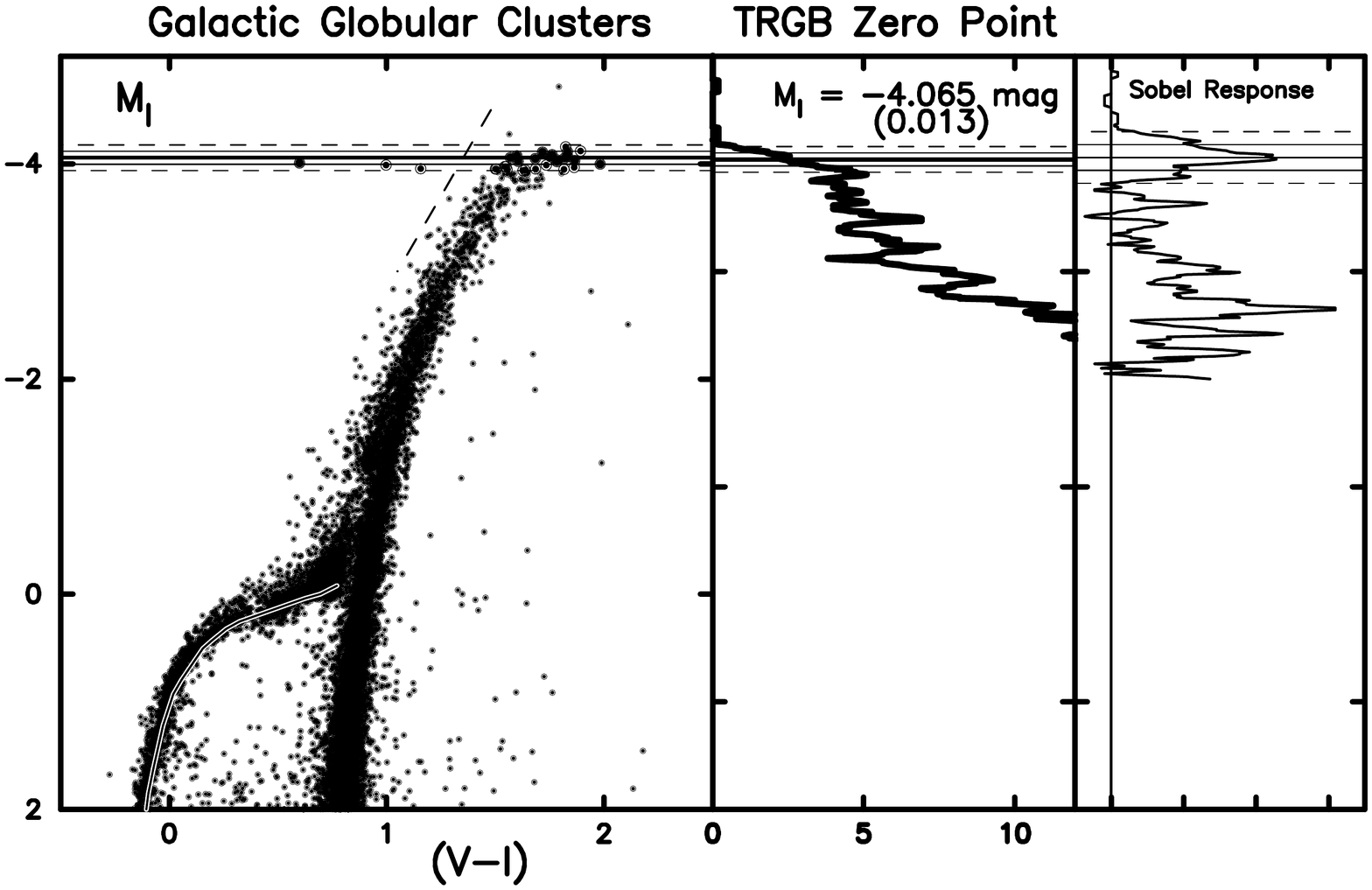}
\caption{\label{fig:Iband_dets} Composite $I$ vs $(V-I)$ CMD, luminosity function, and edge detector response for four bins of equal stellar counts, sorted by increasing metallicity from the top left to the bottom right. We note that these bins are distinct from those used in the calibration procedure, which varied significantly in stellar counts. 
(Left Panels) Optical CMDs for clusters. We apply a loose color--magnitude selection, denoted by the dashed diagonal line, when constructing the CMD and luminosity function, and when measuring the edge response in the center and right-hand panels, respectively. 
(Center) GLOESS-smoothed luminosity function of RGB stars with $\rm M_I > -2.0$~mag, binned into bins of 0.01~mag. (Right) Edge detector response from the Sobel-filtered luminosity function.}
\end{figure*}

\par We apply distance and extinction/reddening corrections using these values when generating the individual cluster CMDs in \figref{all_optical_cmds} and all composite CMDs throughout this work. 

\section{Detecting the TRGB}
\label{sec:measurement}
In the following sections, we utilize the composite cluster sample constructed above to measure the magnitude/slope of the TRGB across five different photometric bandpasses. 

\subsection{Optical: $I$-Band}
We first sought to measure the TRGB in the $I$-band, where the TRGB is expected to be roughly flat across a broad range of photometric color ($ 1.3 \lesssim (V - I) \lesssim 2.2$~mag) based on both stellar evolution modelling and previous empirical studies \citep[e.g.,][and references therein]{freedman_2020}. To do so, we follow the same procedure and utilize the identical code used in \citet{freedman_2019} and \citet{freedman_2020}. In brief, we first bin the $I-$band RGB luminosity function (LF) into bins of 0.01~mag, and count the number of stars falling in each bin. This LF is then smoothed using GLOESS (Gaussian-windowed, Locally-Weighted Scatterplot Smoothing). We then apply a  [−1, 0, +1]  Sobel filter, which acts as an ``edge detector" that responds to discontinuities or steep derivatives in the LF, as expected due to the rapid drop-off in stellar counts at the magnitude of the (flat) $I-$band TRGB.
\par In \figref{Iband_dets}, we present plots depicting four individual measurements of the $I-$band TRGB absolute magnitude using the method described above. The 46-cluster sample is sorted here by increasing metallicity, splitting the composite catalog into four bins of equal numbers of stars. We note that these re-defined metallicity groups are distinct from those used in the calibration procedure (\autoref{sec:calibration}), where the numbers of stars per bin was not constant. 
The $I-$band TRGB absolute magnitude and its statistical uncertainty are shown in each of the individual plots. The range of TRGB magnitudes is only 0.020~mag, and the values for each individual metallicity group agree to within their measured uncertainties. 
\par In \figref{full_composite_Iband}, we present the same plots for the 46-cluster composite. For this full composite, we apply a color selection, indicated by a dashed diagonal line in the right-most panel, in order to remove contaminating sources from our TRGB detection.

\begin{deluxetable}{l c}[h]
\tablecolumns{2}
\tabletypesize{\small}
\tablecaption{\label{tab:optical_results}
$I-$band TRGB detections based on cluster bins of equal counts ($\roughly 90,000$ stars per metallicity bin across all magnitudes), and corresponding statistical uncertainties (before accounting for all other uncertainties). These measurements originate from the detections presented in \figref{Iband_dets}.}
\tablehead{
\colhead{Bin Color Range at Tip}  & \colhead{$M^{I}_{\rm TRGB}$}}
\startdata
$1.35 \lesssim (V-I) \lesssim 1.5$ & -4.045 $\pm 0.012$ (stat) \\
$1.45 \lesssim (V-I) \lesssim 1.55$  & -4.050 $\pm 0.013$ (stat) \\
$1.5 \lesssim (V-I) \lesssim 1.6$  & -4.051 $\pm 0.013$ (stat) \\
$1.6 \lesssim (V-I) \lesssim 1.9$ & -4.065 $\pm 0.013$ (stat) \\
\hline
\textbf{Full Composite:} & -4.056 $\pm 0.012$ (stat) \\
\enddata
\vspace{-3em}
\end{deluxetable}

\subsubsection{Statistical Error on the Mean}

\begin{figure*}
\center
\includegraphics[width=.7\textwidth]{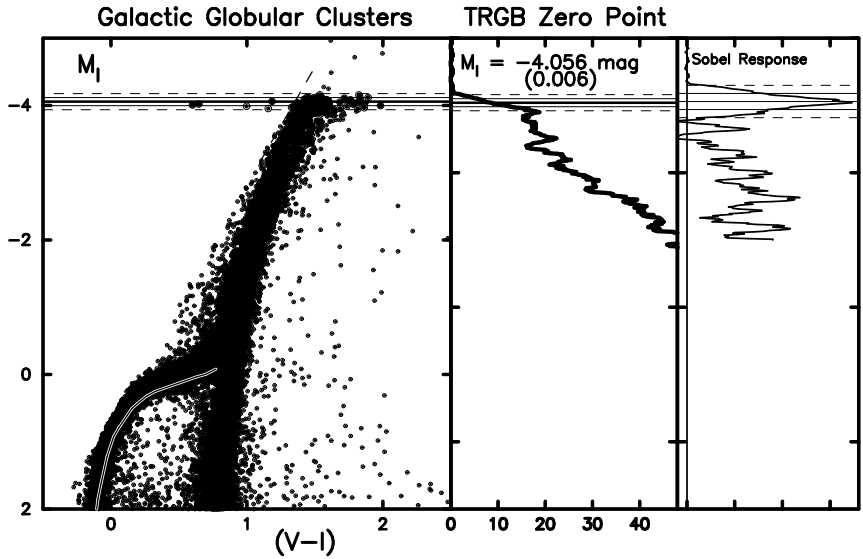}
\caption{Full 46-cluster composite CMDS displaying the upper RGB/TRGB in $B$ vs $(V-I)$, $V$ vs $(V-I)$, and $I$ vs $(V-I)$. We impose a color selection in the $I-$band, removing stars bluer than the dashed diagonal line represented by $M_I = -4.33 (V-I) +1.93$.}
\label{fig:full_composite_Iband}
\end{figure*}

In the I-band CMD (right panel) we include two (thin) horizontal lines to denote one and two sigma above and below the (thick) solid line at the level of our measured TRGB. Contributing to the TRGB detection there are 107 stars within two sigma of the tip. Our technique for aligning the horizontal branches of the clusters contributes to the observed `blurring' of the TRGB, in the plot of the marginalized luminosity function in \figref{full_composite_Iband}. Applying a Sobel filter to the observed luminosity function, we measure the tip at $M_{I}^{TRGB} = -4.056$~mag. Using the statistical error of $\pm$0.039~mag as found in Subsection 4.3 to be representative of the error attributable to each of the N = 4 metallicity bins, we estimated the scatter induced in the TRGB edge to be $\sigma_{stat} = \pm 0.039/\sqrt{(N-1)} = \pm$~0.022~mag, which we adopt as the statistical error on the mean for the I-band TRGB absolute magnitude.

\subsubsection{Adopted Systematic Error}

There are two sources of systematic uncertainty in the TRGB calibration:
(1) The systematic uncertainty in the measured distance modulus to $\omega~Cen$, for which we have adopted the DEB measurement from \citet{thompson_2001}, and
(2) An additional 0.015~mag systematic uncertainty in the DEB distance propagated to the $I-$band apparent modulus measurement from a 0.01~mag uncertainty in the $\omega~Cen$ $E(B-V)$ color excess.
\par Recent work by \citet{Braga_2018} using near infrared observations of RR Lyrae variables
in $\omega~Cen$ have confirmed the DEB distance,  but in addition, their high precision suggests that the quoted systematic uncertainty on the DEB distance from \citet{thompson_2001} may have been overestimated. 
Using a theoretical calibration of the RR Lyrae variables, Braga et al. find true distance moduli for $\omega~Cen$ of 13.674 $\pm$ 0.008 $\pm$ 0.038 mag (statistical error and standard deviation of the median, respectively) using spectroscopic iron abundances, and 13.698 $\pm$ 0.004 $\pm$ 0.048~mag using photometric iron abundances. These are to be compared to the DEB true (geometric) distance modulus of 13.678 $\pm$ 0.11~mag (sys) (\citealt{thompson_2001}) where the largest difference between the three values given above is only 0.02~mag. Given the stability of the estimates of the true distance modulus to $\omega~Cen$ over time, and over methods, (see Figure 23 of \citealt{Braga_2018}) we very conservatively adopt $\pm 0.10$~mag  as the systematic error on the true distance to $\omega~Cen$. This, of course, will soon be superseded with the release of Gaia DR3 (see \secref{discussion}).
Our final adopted TRGB calibration using $\omega~Cen$ is $M_{I}^{TRGB} = -4.06 \pm 0.02 \text{ (stat})  \pm 0.10 \text{ (sys})$~mag. We summarize the quoted errors for this measurement in \tabref{error}.

\begin{deluxetable}{l l l}[h]
\tablecolumns{3}
\tabletypesize{\small}
\tablecaption{\label{tab:error}
Measurement Error Budget}
\tablehead{
\colhead{Source of Uncertainty} & \colhead{$\sigma_{\rm stat}$} & \colhead{$\sigma_{\rm sys}$}}
\startdata
Reddening Alignment & $0.015$ & ~~... \\
Intra-Bin ZAHB Alignment &  $0.03$ & ~~... \\
Inter-Bin ZAHB Alignment &  $0.02$ & ~~... \\
$\omega~Cen$ Distance &  ~~... & $0.10$ \\
$\omega~Cen$ Extinction &  ~~... & $0.015$ \\
\hline 
Quadrature Sum &  $0.039$ & $0.101$ \\
Error on the Mean ($N=4$ ref. bins)&  $0.022$ & ...  \\
\hline
\textbf{Total Uncertainty (mag.)} & $\mathbf{ 0.022}$ & $\mathbf{ 0.101}$ 
\enddata
\tablecomments{TRGB measurement uncertainties due to reddening were propagated from $0.01$ mag uncertainty in $E(B-V)$ to the $I$-band extinction adopting $R_I = 1.485$ when deriving the distance modulus.}
\vspace{-3em}
\end{deluxetable}

\subsection{Optical: $V-$Band}
Because the same stars that define the tip in the $I$-band also define the TRGB across all other wavelengths -- from the optical to the near-infrared -- the procedure for detecting the TRGB in other filter bands is fundamentally simple.  In the $V$ bands, theory predicts (and observations have confirmed) that the TRGB exhibits a positive slope with increasing (redder) stellar color. \citep[e.g.][]{freedman_2020}.
\par As described in detail in \citet{math_paper}, one only needs knowledge of the calibrated run of the TRGB with photometric color and the color-color relations between two bandpasses (for a limited run of spectral types) in order to predict the slope and the zero-point for any other bandpass combination. The application of this type of relation is particularly simple in the case of transforming between the run of the TRGB in $I$ vs. $(V-I)$ to $V$ vs. $(V-I)$: by assuming zero slope for the $I-$band TRGB with respect to $V-I$ color, the $V-$band slope is uniquely determined as $V/(V-I) = 1.00$ . The $V-$band zero point can then be easily determined by adopting a fiducial intrinsic color at which to measure the zero point, and simply adding it to the $I-$band zero point, as $V = I + (V-I)$. Adopting a fiducial measurement color of $(V-I) = 1.8$~mag to match the convention of \citet{freedman_2020}, we thus find a $V-$band zero point of $M_{V} = -2.26$~mag.
\par
\begin{figure*}
\center
\includegraphics[width=.6\textwidth]{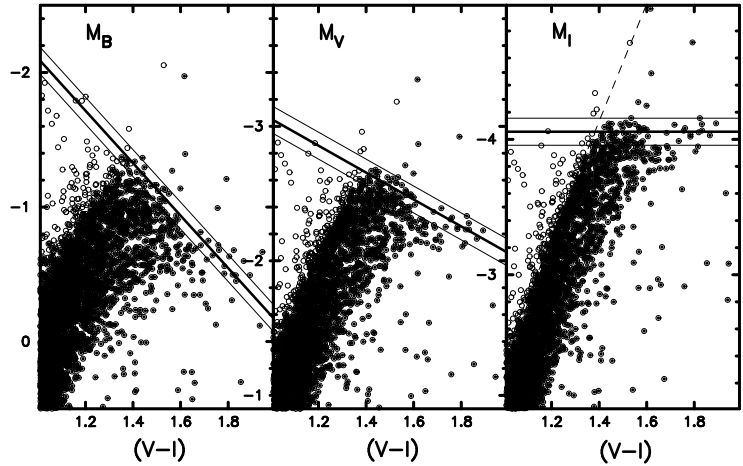}
\caption{Full 46-cluster composite CMDs for all three visual wavelengths. The dark line in each panel corresponds to the derived equations from \secref{measurement}. We note that the $B-$band photometry utilized in this work was found to have significantly larger scatter than the $V$ and $I$ bands, and thus we did not seek to derive a zero point in this filter; however, we include an approximate linear fit to the corresponding CMD in order to display the wavelength-dependent variation of the TRGB across each of the panels.}
\label{fig:comp_tryptch}
\end{figure*}

 To summarize, we find:
\begin{align*}
M_{I}  = -4.06 + 0.0[(V-I) - 1.80] \\  
M_{V}  = -2.26~ + 1.0[(V-I) - 1.80] 
\end{align*}
In \figref{comp_tryptch}, we overplot these derived relations on close-up CMDs of our 46-cluster composite. While we do not calibrate the $B-$band slope and zero point in this work, primarily due to increased scatter in the photometry for this filter, we include an illustrative slope fit to highlight for the reader the wavelength dependence of the TRGB across the three panels.

\subsection{Near Infrared: J,H, and K Bands}
\label{ssec:NIR_Detection}
In order to extend our analysis to measure the TRGB in the near-infrared,  where the TRGB is expected to negatively correlate (decrease in absolute magnitude) with redder intrinsic stellar colors \citep[e.g.][]{madore_2018_nirCalibration, hoyt18}, we undertake a procedure similar to the one described in the previous subsection. However, in doing so, we emphasize two important distinctions from our analysis in the optical bands. 
First, we note that the impact of line-of-sight reddening is greatly diminished in the near-infrared compared to the visual bands, where, for example, the magnitude of extinction in the $K$-band ($A_{K}$) is nearly one tenth of the extinction in the $V-$band ($A_K = .114 A_{V}$; \citealt{C89}). 
Secondly, as introduced in \secref{data}, we now utilize high-precision $JHK$ photometry from 2MASS for stars cross-matched between S19, \Gaia DR2, and S19's catalogs that were identified as likely cluster members in \secref{data}. 
\par In order to derive a TRGB zero point for the $J,K$ bands, we assume the \citet{madore_2018_nirCalibration} $J-$band slope of $J/(J-K) = -0.85$; the $K-$band slope then differs by 1.00, and thus, $K/(J-K) = -1.85$ (see previous subsection and \citealt{math_paper}). These slopes were measured based on TRGB stars in the (low-extinction) halo of the Local Group dwarf galaxy, IC1613. For completeness, the uncertainties on the slopes from that work are $\pm 0.09$ and $\pm 0.19$,  for the $J/(J-K)$ and $K/(J-K)$ slopes, respectively.  
Using these slopes, we then ``rectify" (i.e. flatten) the composite CMDs such that their TRGBs appear flat as a function of intrinsic color (as is naturally the case in the $I-$band).
In doing so, the RGB luminosity functions show the greatest contrast in stellar counts at the TRGB, allowing for the determination of the NIR TRGB zero points through applying the same Sobel filter edge detection as employed in the previous subsections. Similar to the procedure for the $V-$band, we adopt a fiducial color at which to measure the TRGB zero point, namely $(J-K) = 1.00$~mag,\footnote{This color is chosen as it corresponds to the same metallicity as the fiducial color chosen for the $I-$band, $(V-I) = 1.8.$} and measure the $J-$band zero point of $M_{J} = -5.16$~mag, which simultaneously determines the $K-$band zero point to be (by definition) $M_{K} = M_{J} - 1.00 = -6.16$~mag. 
These derived zero points above are based on a large sample of 119 (254) stars within one (two) sigma of the NIR TRGB, reported based on the $J-$band edge detector output.
\par
\begin{figure*}
\center
\includegraphics[width=.4\textwidth]{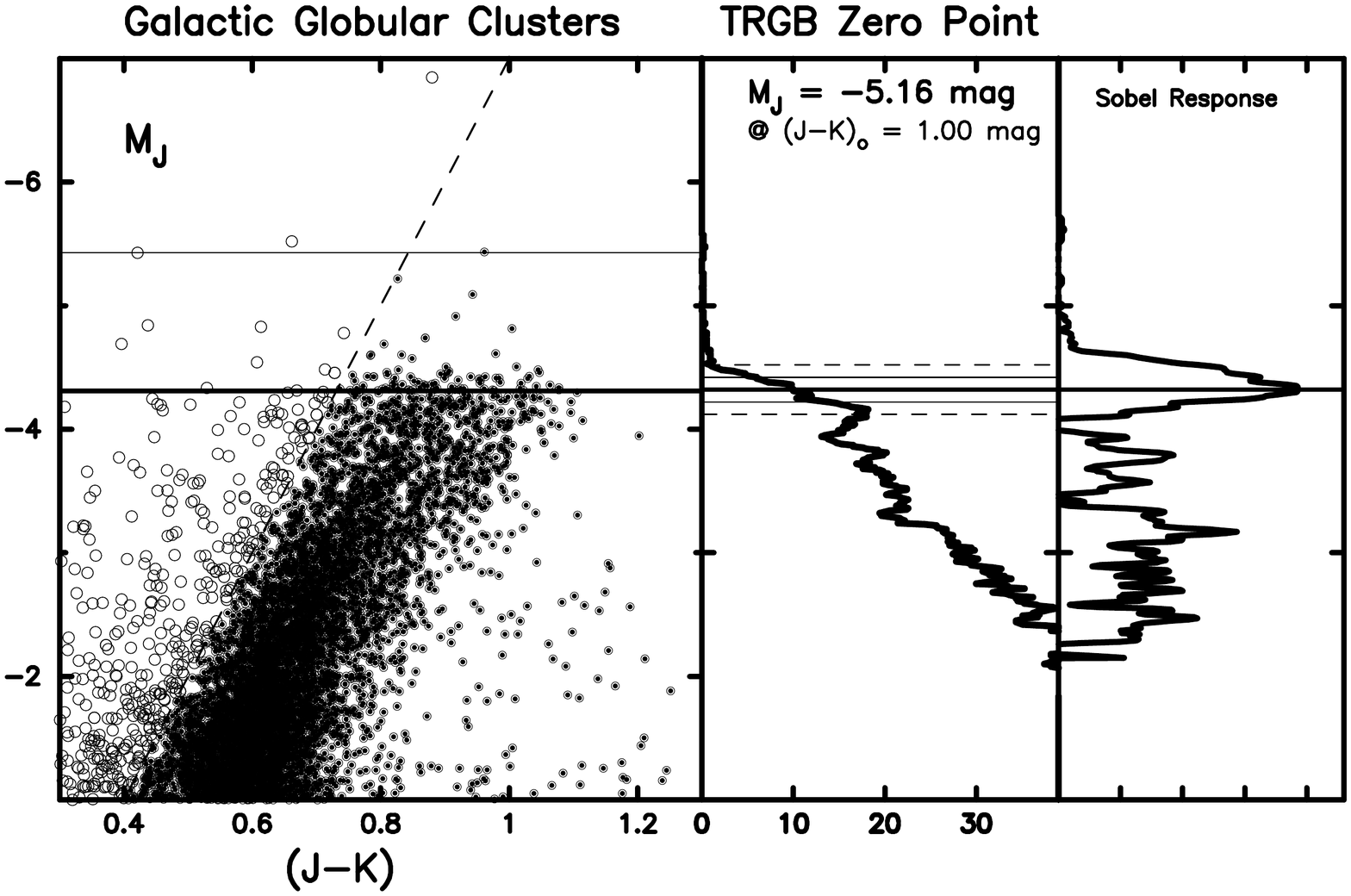}
\includegraphics[width=.4\textwidth]{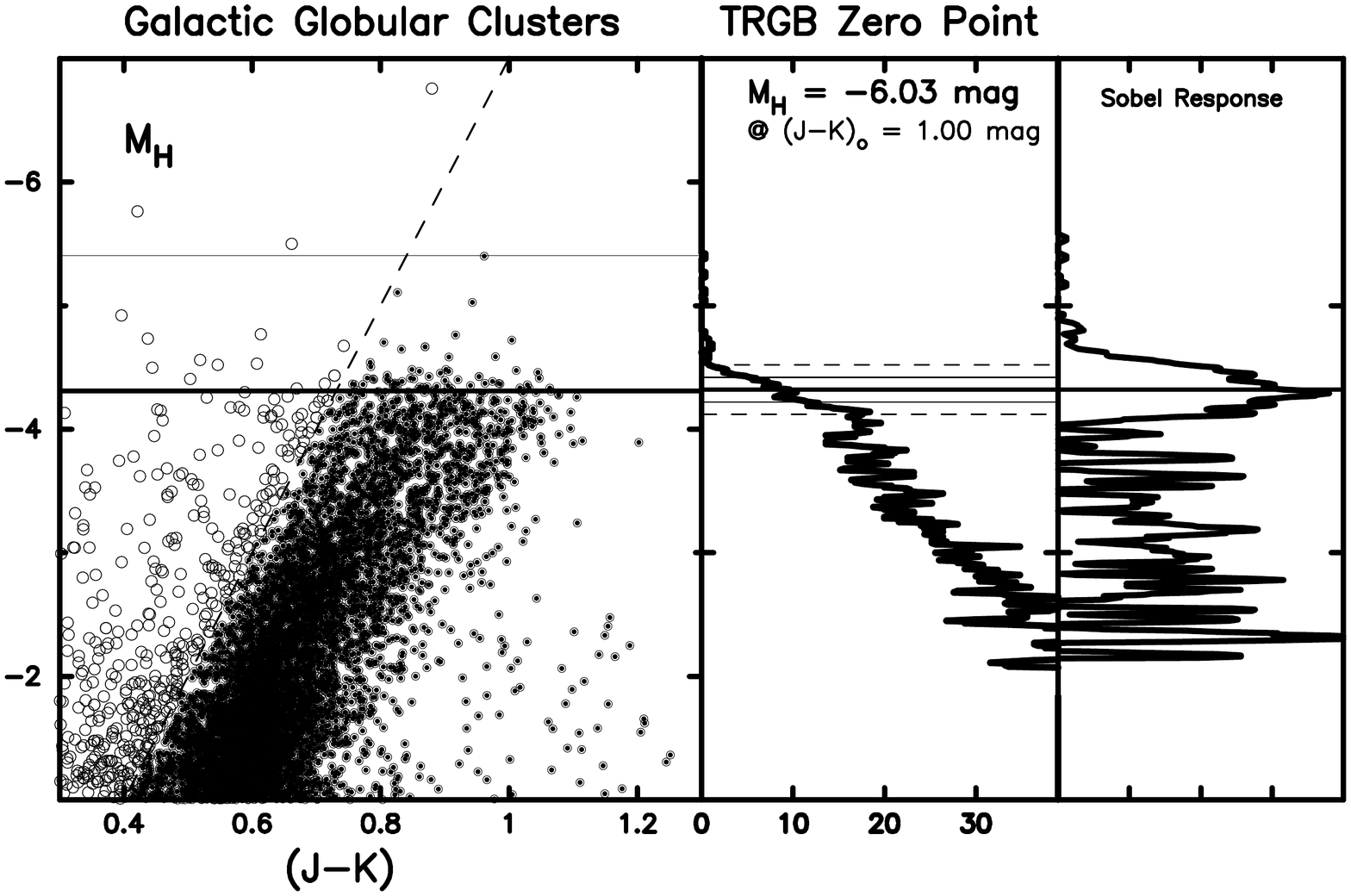}
\includegraphics[width=.4\textwidth]{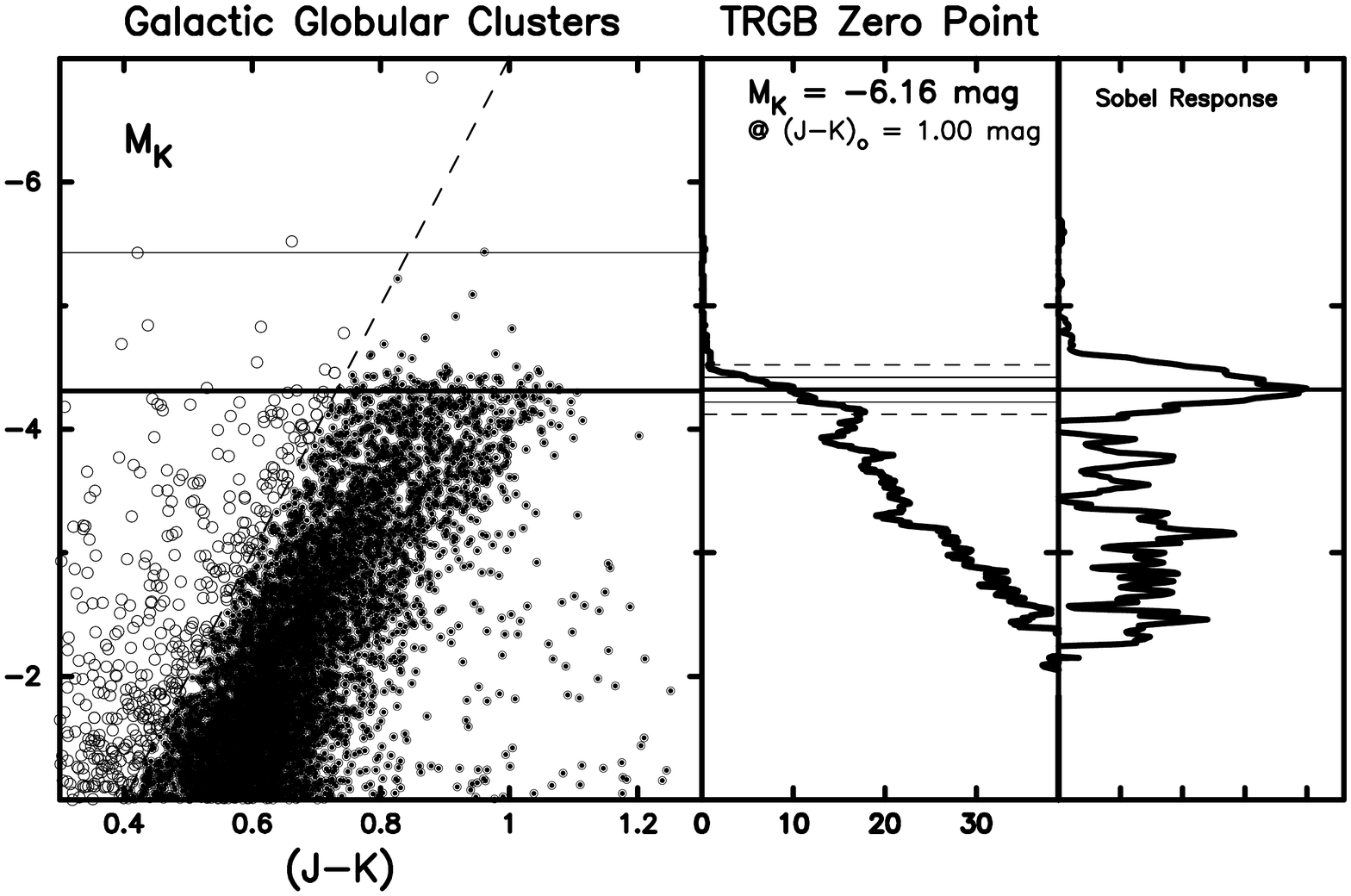}
\caption{\label{fig:all_NIR_rectified} Rectified (flattened) CMDs for the $J,H,K$ bands,  smoothed LFs, and Sobel filter responses for each of the three NIR filter bands. (Left Panels) NIR CMDs for the upper RGB/TRGB of the 46-cluster composite. We apply a color cut indicated by the dashed diagonal line to remove bluer stars that likely belong to the asymptotic giant branch and other contaminant stars. The upper horizontal line in each plot at $M = -5.5$~mag corresponds to the brightest magnitude included in the LF for each band after rectification. The other (lower) horizontal line corresponds to the rectified TRGB edge detection magnitude. (Center Panels) Smoothed luminosity function for each of the three composite CMDs. The magnitude of greatest edge response is indicated with a dark line, with the one and two sigma error bars drawn as dashed lines. As expected mathematically, the $J-$ and $K-$band zero points differ by exactly 1.00~mag. (Right Panels) Sobel filter edge response }
\end{figure*}

 For the $H-$band, we re-fit the TRGB slope directly via simple linear regression over all stars within two sigma of the rectified $J-$band edge detection, deriving a slope of $H/(J-K) = -1.72$. Then, holding the $H-$band zero point to be the same as the $J,K$ zero point at $(J-K) = 0.00$~mag, we run the Sobel filter over the rectified $H-$band data, deriving a zero point of $M_{H} = -6.03$~mag.
\figref{all_NIR_rectified} depicts rectified CMDs, smoothed LFs, and Sobel filter responses for each of the three NIR filter bands. In the left panels of each subplot, we include a CMD of the upper RGB and TRGB for the 46-cluster composite, where the color-dependence of the TRGB has been flattened. The dashed, diagonal line represents a color cut applied before the LF is constructed and the Sobel filter edge detector is run.

\par In summary, we find: 
\begin{align*}
M_{J} = -5.16 - 0.85  [(J-K) - 1.00] \\
M_{H} = -6.03 - 1.72  [(J-K) - 1.00] \\
M_{K} = -6.16 - 1.85  [(J-K) - 1.00] 
\setlength{\belowdisplayskip}{0pt}%
\vspace{-5.0em}
\end{align*}\normalsize
where we have re-derived the zero points and $H-$band slope by assuming the $J,K$ TRGB slopes from \citep{madore_2018_nirCalibration}.

\par
\begin{figure*}
\center
\includegraphics[width=.8\textwidth]{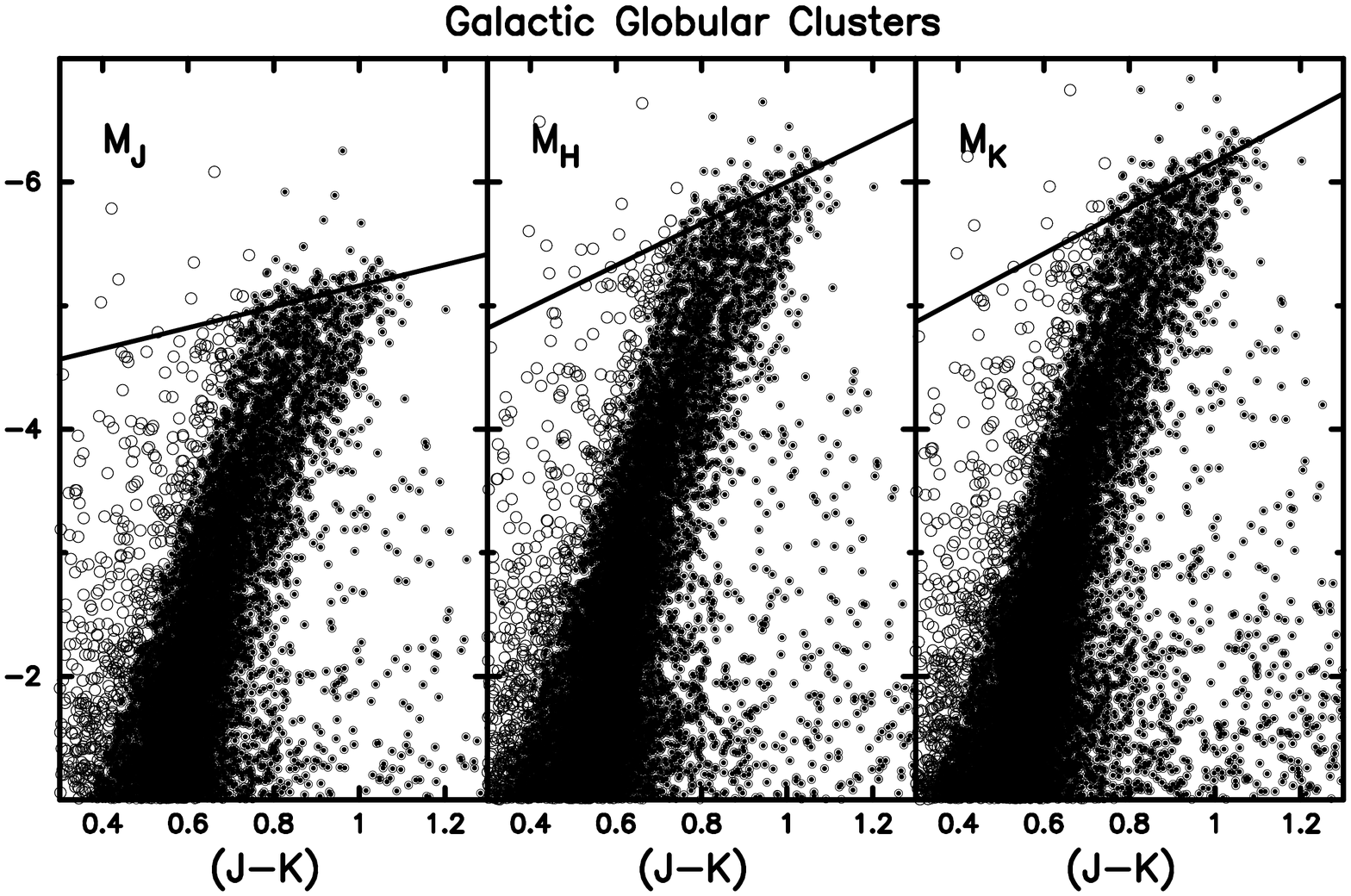}
\caption{Full 46-cluster composite CMDs based on the 2MASS $J,H,K$ photometry. The diagonal line corresponds in each panel represent the calibration equations presented in Section \ref{ssec:NIR_Detection}. Unfilled circles represent those excluded from when deriving the zero point due a color cut. 
}
\label{fig:JHK_slopes}
\end{figure*}

 In \figref{JHK_slopes}, we present $J$ vs $(J-K)$, $H$ vs. $(J-K)$, and $K$ vs $(J-K)$ composite CMDs depicting the upper RGB and TRGB for the 46-cluster sample using the 2MASS photometry. In each panel, we plot a diagonal line indicating the TRGB calibration described by each of the above equations.

\section{Discussion}

\label{sec:discussion}
\subsection{Comparison with Prior Measurements}
In \tabref{comparison} we summarize our measurements of the TRGB absolute magnitude and slope. We place these results in the context of calibrations derived from a variety of physical systems, including other studies of Galactic globular clusters (\citealt{bellazzini_2004}, \citealt{freedman_2020}), the LMC (\citealt{hoyt18,freedman_2020, Gorski_2018}), and the maser-hosting galaxy NGC~4258 \citep{maser_trgb}. We find excellent agreement with the body of existing literature measurements, with our measurement lying within one sigma of nearly all tabulated measurements across all wavelengths considered in this work.

\begin{deluxetable*}{l c c c}
\tablecolumns{4}
\tabletypesize{\small}
\tablecaption{\label{tab:comparison}Comparison of TRGB calibrations}
\tablehead{
\colhead{Filter} &  \colhead{Zeropoint} & \colhead{Reference} & \colhead{Calibration Technique}}
\startdata
$I$&  $-4.05 \pm 0.12$ & \citet{bellazzini_2004} & DEB distances to 47 Tuc, $\omega$ Cen (averaged)\\
$I$&  $-4.056 \pm 0.053 \text{ (stat)} \pm 0.080 \text{ (sys)}$ & \citet{freedman_2020} & 12 GCs with literature distances + DEB distance to 47 Tuc\\
$I$&  $-4.047 \pm 0.022 \text{ (stat)} \pm 0.039 \text{ (sys)}$ & \citet{freedman_2020} & 18 DEBs in LMC \\
$F814W$&  $-4.051 \pm 0.027 \text{ (stat)} \pm 0.045 \text{ (sys)}$ & \citet{maser_trgb} & Maser in NGC4258 \\
$I$ & \textbf{$-4.056 \pm 0.022 \text{ (stat)} \pm 0.1 \text{ (sys)}$} & \textbf{This Work} & \textbf{46 GCs - Cluster ZAHBs and DEB distance to $\omega$ Cen}\\
\hline 
$V$&  $-2.254 \pm 0.022 \text{ (stat)} \pm 0.039 \text{ (sys)}$ & \citet{freedman_2020} & DEBs in LMC \\
$V$&  $-2.256 \pm 0.053 \text{ (stat)} \pm 0.080 \text{ (sys)}$ & \citet{freedman_2020} & 12 GCs with literature distances + DEB distance to 47 Tuc\\
$V$&  $-2.251 \pm 0.027 \text{ (stat)} \pm 0.045 \text{ (sys)}$ & \citet{maser_trgb} & Maser in NGC4258 \\
$V$ & \textbf{$-2.26 \pm 0.022 \text{ (stat)} \pm 0.1 \text{ (sys)}$} & \textbf{This Work} & \textbf{46 GCs - Cluster ZAHBs and DEB distance to $\omega$ Cen}\\
\hline
$J$&  $-5.20 \pm 0.01 \text{ (stat)} \pm 0.05 \text{ (sys)}$ & \citet{hoyt18} & 8 DEBs in LMC \\
$J$&  $-5.241 \pm 0.007 \text{ (stat)} \pm 0.050 \text{ (sys)}$ & \citet{Gorski_2018} & 8 LMC DEBs; 4 SMC DEBs \\
$J$&  $-5.14 \pm 0.022 \text{ (stat)} \pm 0.039 \text{ (sys)}$ & \citet{freedman_2020} & 18 DEBs in LMC \\
$J$ & \textbf{$-5.16 \pm 0.022 \text{ (stat)} \pm 0.1 \text{ (sys)}$} & \textbf{This Work} & \textbf{46 GCs - Cluster ZAHBs and DEB distance to $\omega$ Cen}\\
\hline
$H$&  $-5.98 \pm 0.01 \text{ (stat)} \pm 0.05 \text{ (sys)}$ & \citet{hoyt18} & 8 DEBs in LMC \\
$H$&  $-6.072 \pm 0.008 \text{ (stat)} \pm 0.048 \text{ (sys)}$ & \citet{Gorski_2018} & 8 LMC DEBs; 4 SMC DEBs \\
$H$ & \textbf{$-6.03 \pm 0.022 \text{ (stat)} \pm 0.1 \text{ (sys)}$} & \textbf{This Work} & \textbf{46 GCs - Cluster ZAHBs and DEB distance to $\omega$ Cen}\\
\hline
$K$&  $-6.16 \pm 0.01 \text{ (stat)} \pm 0.05 \text{ (sys)}$ & \citet{hoyt18} & 8 DEBs in LMC \\
$K$&  $-6.264 \pm 0.008 \text{ (stat)} \pm 0.048 \text{ (sys)}$ & \citet{Gorski_2018} & 8 LMC DEBs; 4 SMC DEBs \\
$K$ & \textbf{$-6.16 \pm 0.022 \text{ (stat)} \pm 0.1 \text{ (sys)}$} & \textbf{This Work} & \textbf{46 GCs - Cluster ZAHBs and DEB distance} to $\omega$ Cen\\
\enddata
\tablecomments{The \citet{maser_trgb} $V-$band TRGB measurement listed here is converted from that work's $F814W-$band measurement following the same technique utilized in \secref{calibration}, marginalizing over the difference between the $F814W$ and Johnson-Cousins $I$ filters (and thus should be considered approximate). The \citet{hoyt18} zero-points are rescaled here to the reddening value determined in \citet{freedman_2020}. The statistical uncertainties for $J,H,K$ from \citet{Gorski_2018} are taken to be the errors on their fits to the TRGB magnitude relations (their Table 3), and the systematic error is taken to be the quadrature sum of the uncertainty in their adopted LMC distance modulus and the typical uncertainty in their absolute reddening zero-point.} 
\end{deluxetable*}

\subsection{Future Prospects with \Gaia DR3}
While our analysis did not utilize parallaxes to select cluster member stars or to set our absolute zero-point, the upcoming \Gaia Early Data Release 3 (EDR3; expected December 2020) and planned full Data Release 3 (DR3; currently planned for 2022) will offer the possibility of calibrating the TRGB based solely on geometrically-defined distances to these clusters. Such an advancement will be enabled both by corrections to the underlying astrometric solution (currently known to exhibit a systematic parallax bias with respect to background quasars; \citealt{2019MNRAS.487.3568S}), and also significantly improved astrometric systematics. Because the error in our TRGB measurements is dominated by the systematic error on the single \citet{thompson_2001} DEB distance, the use of parallax-based geometric distances from EDR3 to each of the 46 clusters considered here have the potential to improve  the precision and accuracy of this cluster-based calibration of the TRGB luminosity by an order of magnitude. Additionally, EDR3's expected factor-of-two improvements in proper motion uncertainties, 20$\%$ improvements in parallax uncertainties, and significant increase in the number of stellar sources with full 5-parameter astrometric measurements will make it possible to increase the purity and completeness of cluster membership catalogs -- for example, by utilizing parallaxes as a parameter in mixture model classifications (\secref{data}), capitalizing on more precise proper motion measurements of stars located near the cores of clusters, and by reaping the benefits of increasingly confident probabilistic estimates of a star's membership likelihood due to the refined proper motion measurements. 
\section{Conclusion}
 Using \Gaia DR2, we have applied Gaussian Mixture Model clustering to precisely distinguish member stars for 46 Galactic globular clusters, and crossmatched with homogeneous photometric datasets in the optical and near-infrared bandpasses to build a multi-band catalog of cluster data spanning a wide range of metallicity. Using the zero-age horizontal branch, we brought these 46 clusters onto a consistent relative calibration, before re-zeroing the ensuing composite to a detached eclipsing binary (geometric) distance measurement to $\omega~Cen$. We then utilized this uniformly-calibrated cluster composite to measure the zero point of the TRGB across five optical and near-infrared wavelengths, finding excellent agreement with previous calibrations. 
\par Our results provide an independent check on the application of the TRGB to nearby galaxies, reinforcing the validity of TRGB-based rungs of the cosmological distance ladder. The methods and calibrations outlined in this paper will be greatly improved with the advent of \Gaia DR3, which will bring reliable, self-consistent parallax distances to these clusters,  allowing for greatly diminished systematic uncertainties associated with each measurement and the overall absolute cluster zero point, along with improvements to the astrometry underlying the cluster member star selection process.

\section{Acknowledgments}
We thank the {\it Observatories of the Carnegie Institution for
Science} and the {\it University of Chicago} for their support of our long-term research
into the calibration and determination of the expansion rate of the Universe.

\par This work has made use of data from the European Space Agency (ESA) mission {\it Gaia} (\url{https://www.cosmos.esa.int/gaia}), processed by the {\it Gaia} Data Processing and Analysis Consortium (DPAC, \url{https://www.cosmos.esa.int/web/gaia/dpac/consortium}).
Funding for the DPAC has been provided by national institutions, in particular the institutions participating in the {\it Gaia} Multilateral Agreement.
\par This research made use of the cross-match service provided by CDS, Strasbourg ([http://cdsxmatch.u-strasbg.fr/]) 

\facility{\Gaia.}
\software{\code{astropy} \citep{astropy:2013,astropy:2018}, \code{Matplotlib} \citep{Hunter:2007}, \code{numpy} \citep{numpy:2011}, \code{scikit-learn} \citep{sklearn}}

\bibliography{main}

\begin{thebibliography}{}
\expandafter\ifx\csname natexlab\endcsname\relax\def\natexlab#1{#1}\fi
\providecommand{\url}[1]{\href{#1}{#1}}

\bibitem[{{Astropy Collaboration} {et~al.}(2013){Astropy Collaboration},
  {Robitaille}, {Tollerud}, {Greenfield}, {Droettboom}, {Bray}, {Aldcroft},
  {Davis}, {Ginsburg}, {Price-Whelan}, {Kerzendorf}, {Conley}, {Crighton},
  {Barbary}, {Muna}, {Ferguson}, {Grollier}, {Parikh}, {Nair}, {Unther},
  {Deil}, {Woillez}, {Conseil}, {Kramer}, {Turner}, {Singer}, {Fox}, {Weaver},
  {Zabalza}, {Edwards}, {Azalee Bostroem}, {Burke}, {Casey}, {Crawford},
  {Dencheva}, {Ely}, {Jenness}, {Labrie}, {Lim}, {Pierfederici}, {Pontzen},
  {Ptak}, {Refsdal}, {Servillat}, \& {Streicher}}]{astropy:2013}
{Astropy Collaboration}, {Robitaille}, T.~P., {Tollerud}, E.~J., {et~al.} 2013,
  \aap, 558, A33

\bibitem[{{Baade}(1944)}]{baade_1944}
{Baade}, W. 1944, \apj, 100, 137

\bibitem[{{Bellazzini} {et~al.}(2001){Bellazzini}, {Ferraro}, \&
  {Pancino}}]{bellazzini_2001}
{Bellazzini}, M., {Ferraro}, F.~R., \& {Pancino}, E. 2001, \apj, 556, 635

\bibitem[{{Bellazzini} {et~al.}(2004){Bellazzini}, {Ferraro}, {Sollima},
  {Pancino}, \& {Origlia}}]{bellazzini_2004}
{Bellazzini}, M., {Ferraro}, F.~R., {Sollima}, A., {Pancino}, E., \& {Origlia},
  L. 2004, \aap, 424, 199

\bibitem[{Braga {et~al.}(2018)Braga, Stetson, Bono, Dall'Ora, Ferraro,
  Fiorentino, Iannicola, Marconi, Marengo, Monson, Neeley, Persson, Beaton,
  Buonanno, Calamida, Castellani, Carlo, Fabrizio, Freedman, Inno, Madore,
  Magurno, Marchetti, Marinoni, Marrese, Matsunaga, Minniti, Monelli, Nonino,
  Piersimoni, Pietrinferni, Prada-Moroni, Pulone, Stellingwerf, Tognelli,
  Walker, Valenti, \& Zoccali}]{Braga_2018}
Braga, V.~F., Stetson, P.~B., Bono, G., {et~al.} 2018, The Astronomical
  Journal, 155, 137.
\newblock \url{https://doi.org/10.3847%2F1538-3881%2Faaadab}

\bibitem[{Bustos Fierro \& Calderón(2019)}]{Fierro:19}
Bustos Fierro, I.~H., \& Calderón, J.~H. 2019, Monthly Notices of the Royal
  Astronomical Society, 488, 3024.
\newblock \url{https://doi.org/10.1093/mnras/stz1879}

\bibitem[{{Cardelli} {et~al.}(1989){Cardelli}, {Clayton}, \& {Mathis}}]{C89}
{Cardelli}, J.~A., {Clayton}, G.~C., \& {Mathis}, J.~S. 1989, \apj, 345, 245

\bibitem[{{Da Costa} \& {Armandroff}(1990)}]{da_costa_armandroff_1990}
{Da Costa}, G.~S., \& {Armandroff}, T.~E. 1990, \aj, 100, 162

\bibitem[{{Efstathiou}(2020)}]{efstathiou_2020}
{Efstathiou}, G. 2020, arXiv e-prints, arXiv:2007.10716

\bibitem[{{Ferraro} {et~al.}(1999){Ferraro}, {Messineo}, {Fusi Pecci}, {de
  Palo}, {Straniero}, {Chieffi}, \& {Limongi}}]{ferraro_1999}
{Ferraro}, F.~R., {Messineo}, M., {Fusi Pecci}, F., {et~al.} 1999, \aj, 118,
  1738

\bibitem[{{Freedman}(2017)}]{freedman_2017}
{Freedman}, W.~L. 2017, Nature Astronomy, 1, 0121

\bibitem[{{Freedman} {et~al.}(2001){Freedman}, {Madore}, {Gibson}, {Ferrarese},
  {Kelson}, {Sakai}, {Mould}, {Kennicutt}, {Ford}, {Graham}, {Huchra},
  {Hughes}, {Illingworth}, {Macri}, \& {Stetson}}]{freedman_2001}
{Freedman}, W.~L., {Madore}, B.~F., {Gibson}, B.~K., {et~al.} 2001, \apj, 553,
  47

\bibitem[{{Freedman} {et~al.}(2019){Freedman}, {Madore}, {Hatt}, {Hoyt},
  {Jang}, {Beaton}, {Burns}, {Lee}, {Monson}, {Neeley}, {Phillips}, {Rich}, \&
  {Seibert}}]{freedman_2019}
{Freedman}, W.~L., {Madore}, B.~F., {Hatt}, D., {et~al.} 2019, \apj, 882, 34

\bibitem[{{Freedman} {et~al.}(2020){Freedman}, {Madore}, {Hoyt}, {Jang},
  {Beaton}, {Lee}, {Monson}, {Neeley}, \& {Rich}}]{freedman_2020}
{Freedman}, W.~L., {Madore}, B.~F., {Hoyt}, T., {et~al.} 2020, \apj, 891, 57

\bibitem[{{Gaia Collaboration} {et~al.}(2018){Gaia Collaboration}, {Helmi, A.},
  {van Leeuwen, F.}, {McMillan, P. J.}, {Massari, D.}, {Antoja, T.}, {Robin, A.
  C.}, {Lindegren, L.}, {Bastian, U.}, {Arenou, F.}, {Babusiaux, C.},
  {Biermann, M.}, {Breddels, M. A.}, {Hobbs, D.}, {Jordi, C.}, {Pancino, E.},
  {Reyl\'e, C.}, {Veljanoski, J.}, {Brown, A. G. A.}, {Vallenari, A.}, {Prusti,
  T.}, {de Bruijne, J. H. J.}, {Bailer-Jones, C. A. L.}, {Evans, D. W.}, {Eyer,
  L.}, {Jansen, F.}, {Klioner, S. A.}, {Lammers, U.}, {Luri, X.}, {Mignard,
  F.}, {Panem, C.}, {Pourbaix, D.}, {Randich, S.}, {Sartoretti, P.}, {Siddiqui,
  H. I.}, {Soubiran, C.}, {Walton, N. A.}, {Cropper, M.}, {Drimmel, R.}, {Katz,
  D.}, {Lattanzi, M. G.}, {Bakker, J.}, {Cacciari, C.}, {Casta\~neda, J.},
  {Chaoul, L.}, {Cheek, N.}, {De Angeli, F.}, {Fabricius, C.}, {Guerra, R.},
  {Holl, B.}, {Masana, E.}, {Messineo, R.}, {Mowlavi, N.}, {Nienartowicz, K.},
  {Panuzzo, P.}, {Portell, J.}, {Riello, M.}, {Seabroke, G. M.}, {Tanga, P.},
  {Th\'evenin, F.}, {Gracia-Abril, G.}, {Comoretto, G.}, {Garcia-Reinaldos,
  M.}, {Teyssier, D.}, {Altmann, M.}, {Andrae, R.}, {Audard, M.},
  {Bellas-Velidis, I.}, {Benson, K.}, {Berthier, J.}, {Blomme, R.}, {Burgess,
  P.}, {Busso, G.}, {Carry, B.}, {Cellino, A.}, {Clementini, G.}, {Clotet, M.},
  {Creevey, O.}, {Davidson, M.}, {De Ridder, J.}, {Delchambre, L.},
  {Dell\'{}Oro, A.}, {Ducourant, C.}, {Fern\'andez-Hern\'andez, J.},
  {Fouesneau, M.}, {Fr\'emat, Y.}, {Galluccio, L.}, {Garc\'{\i}a-Torres, M.},
  {Gonz\'alez-N\'u\~nez, J.}, {Gonz\'alez-Vidal, J. J.}, {Gosset, E.}, {Guy, L.
  P.}, {Halbwachs, J.-L.}, {Hambly, N. C.}, {Harrison, D. L.}, {Hern\'andez,
  J.}, {Hestroffer, D.}, {Hodgkin, S. T.}, {Hutton, A.}, {Jasniewicz, G.},
  {Jean-Antoine-Piccolo, A.}, {Jordan, S.}, {Korn, A. J.}, {Krone-Martins, A.},
  {Lanzafame, A. C.}, {Lebzelter, T.}, {L\"offler, W.}, {Manteiga, M.},
  {Marrese, P. M.}, {Mart\'{\i}n-Fleitas, J. M.}, {Moitinho, A.}, {Mora, A.},
  {Muinonen, K.}, {Osinde, J.}, {Pauwels, T.}, {Petit, J.-M.}, {Recio-Blanco,
  A.}, {Richards, P. J.}, {Rimoldini, L.}, {Sarro, L. M.}, {Siopis, C.},
  {Smith, M.}, {Sozzetti, A.}, {S\"uveges, M.}, {Torra, J.}, {van Reeven, W.},
  {Abbas, U.}, {Abreu Aramburu, A.}, {Accart, S.}, {Aerts, C.}, {Altavilla,
  G.}, {\'Alvarez, M. A.}, {Alvarez, R.}, {Alves, J.}, {Anderson, R. I.},
  {Andrei, A. H.}, {Anglada Varela, E.}, {Antiche, E.}, {Arcay, B.},
  {Astraatmadja, T. L.}, {Bach, N.}, {Baker, S. G.}, {Balaguer-N\'u\~nez, L.},
  {Balm, P.}, {Barache, C.}, {Barata, C.}, {Barbato, D.}, {Barblan, F.},
  {Barklem, P. S.}, {Barrado, D.}, {Barros, M.}, {Barstow, M. A.},
  {Bartholom\'e Mu\~noz, S.}, {Bassilana, J.-L.}, {Becciani, U.}, {Bellazzini,
  M.}, {Berihuete, A.}, {Bertone, S.}, {Bianchi, L.}, {Bienaym\'e, O.},
  {Blanco-Cuaresma, S.}, {Boch, T.}, {Boeche, C.}, {Bombrun, A.}, {Borrachero,
  R.}, {Bossini, D.}, {Bouquillon, S.}, {Bourda, G.}, {Bragaglia, A.},
  {Bramante, L.}, {Bressan, A.}, {Brouillet, N.}, {Br\"usemeister, T.},
  {Brugaletta, E.}, {Bucciarelli, B.}, {Burlacu, A.}, {Busonero, D.},
  {Butkevich, A. G.}, {Buzzi, R.}, {Caffau, E.}, {Cancelliere, R.},
  {Cannizzaro, G.}, {Cantat-Gaudin, T.}, {Carballo, R.}, {Carlucci, T.},
  {Carrasco, J. M.}, {Casamiquela, L.}, {Castellani, M.}, {Castro-Ginard, A.},
  {Charlot, P.}, {Chemin, L.}, {Chiavassa, A.}, {Cocozza, G.}, {Costigan, G.},
  {Cowell, S.}, {Crifo, F.}, {Crosta, M.}, {Crowley, C.}, {Cuypers, J.},
  {Dafonte, C.}, {Damerdji, Y.}, {Dapergolas, A.}, {David, P.}, {David, M.},
  {de Laverny, P.}, {De Luise, F.}, {De March, R.}, {de Martino, D.}, {de
  Souza, R.}, {de Torres, A.}, {Debosscher, J.}, {del Pozo, E.}, {Delbo, M.},
  {Delgado, A.}, {Delgado, H. E.}, {Di Matteo, P.}, {Diakite, S.}, {Diener,
  C.}, {Distefano, E.}, {Dolding, C.}, {Drazinos, P.}, {Dur\'an, J.},
  {Edvardsson, B.}, {Enke, H.}, {Eriksson, K.}, {Esquej, P.}, {Eynard Bontemps,
  G.}, {Fabre, C.}, {Fabrizio, M.}, {Faigler, S.}, {Falc\~ao, A. J.}, {Farr\`as
  Casas, M.}, {Federici, L.}, {Fedorets, G.}, {Fernique, P.}, {Figueras, F.},
  {Filippi, F.}, {Findeisen, K.}, {Fonti, A.}, {Fraile, E.}, {Fraser, M.},
  {Fr\'ezouls, B.}, {Gai, M.}, {Galleti, S.}, {Garabato, D.},
  {Garc\'{\i}a-Sedano, F.}, {Garofalo, A.}, {Garralda, N.}, {Gavel, A.},
  {Gavras, P.}, {Gerssen, J.}, {Geyer, R.}, {Giacobbe, P.}, {Gilmore, G.},
  {Girona, S.}, {Giuffrida, G.}, {Glass, F.}, {Gomes, M.}, {Granvik, M.},
  {Gueguen, A.}, {Guerrier, A.}, {Guiraud, J.}, {Guti\'errez-S\'anchez, R.},
  {Haigron, R.}, {Hatzidimitriou, D.}, {Hauser, M.}, {Haywood, M.}, {Heiter,
  U.}, {Heu, J.}, {Hilger, T.}, {Hofmann, W.}, {Holland, G.}, {Huckle, H. E.},
  {Hypki, A.}, {Icardi, V.}, {Jan\ss{}en, K.}, {Jevardat de Fombelle, G.},
  {Jonker, P. G.}, {Juh\'asz, \'A. L.}, {Julbe, F.}, {Karampelas, A.}, {Kewley,
  A.}, {Klar, J.}, {Kochoska, A.}, {Kohley, R.}, {Kolenberg, K.}, {Kontizas,
  M.}, {Kontizas, E.}, {Koposov, S. E.}, {Kordopatis, G.},
  {Kostrzewa-Rutkowska, Z.}, {Koubsky, P.}, {Lambert, S.}, {Lanza, A. F.},
  {Lasne, Y.}, {Lavigne, J.-B.}, {Le Fustec, Y.}, {Le Poncin-Lafitte, C.},
  {Lebreton, Y.}, {Leccia, S.}, {Leclerc, N.}, {Lecoeur-Taibi, I.}, {Lenhardt,
  H.}, {Leroux, F.}, {Liao, S.}, {Licata, E.}, {Lindstr\o{}m, H. E. P.},
  {Lister, T. A.}, {Livanou, E.}, {Lobel, A.}, {L\'opez, M.}, {Managau, S.},
  {Mann, R. G.}, {Mantelet, G.}, {Marchal, O.}, {Marchant, J. M.}, {Marconi,
  M.}, {Marinoni, S.}, {Marschalk\'o, G.}, {Marshall, D. J.}, {Martino, M.},
  {Marton, G.}, {Mary, N.}, {Matijevic, G.}, {Mazeh, T.}, {Messina, S.},
  {Michalik, D.}, {Millar, N. R.}, {Molina, D.}, {Molinaro, R.}, {Moln\'ar,
  L.}, {Montegriffo, P.}, {Mor, R.}, {Morbidelli, R.}, {Morel, T.}, {Morris,
  D.}, {Mulone, A. F.}, {Muraveva, T.}, {Musella, I.}, {Nelemans, G.},
  {Nicastro, L.}, {Noval, L.}, {O\'{}Mullane, W.}, {Ord\'enovic, C.},
  {Ord\'o\~nez-Blanco, D.}, {Osborne, P.}, {Pagani, C.}, {Pagano, I.},
  {Pailler, F.}, {Palacin, H.}, {Palaversa, L.}, {Panahi, A.}, {Pawlak, M.},
  {Piersimoni, A. M.}, {Pineau, F.-X.}, {Plachy, E.}, {Plum, G.}, {Poggio, E.},
  {Poujoulet, E.}, {Prsa, A.}, {Pulone, L.}, {Racero, E.}, {Ragaini, S.},
  {Rambaux, N.}, {Ramos-Lerate, M.}, {Regibo, S.}, {Riclet, F.}, {Ripepi, V.},
  {Riva, A.}, {Rivard, A.}, {Rixon, G.}, {Roegiers, T.}, {Roelens, M.},
  {Romero-G\'omez, M.}, {Rowell, N.}, {Royer, F.}, {Ruiz-Dern, L.}, {Sadowski,
  G.}, {Sagrist\`a Sell\'es, T.}, {Sahlmann, J.}, {Salgado, J.}, {Salguero,
  E.}, {Sanna, N.}, {Santana-Ros, T.}, {Sarasso, M.}, {Savietto, H.},
  {Schultheis, M.}, {Sciacca, E.}, {Segol, M.}, {Segovia, J. C.}, {S\'egransan,
  D.}, {Shih, I-C.}, {Siltala, L.}, {Silva, A. F.}, {Smart, R. L.}, {Smith, K.
  W.}, {Solano, E.}, {Solitro, F.}, {Sordo, R.}, {Soria Nieto, S.}, {Souchay,
  J.}, {Spagna, A.}, {Spoto, F.}, {Stampa, U.}, {Steele, I. A.},
  {Steidelm\"uller, H.}, {Stephenson, C. A.}, {Stoev, H.}, {Suess, F. F.},
  {Surdej, J.}, {Szabados, L.}, {Szegedi-Elek, E.}, {Tapiador, D.}, {Taris,
  F.}, {Tauran, G.}, {Taylor, M. B.}, {Teixeira, R.}, {Terrett, D.},
  {Teyssandier, P.}, {Thuillot, W.}, {Titarenko, A.}, {Torra Clotet, F.},
  {Turon, C.}, {Ulla, A.}, {Utrilla, E.}, {Uzzi, S.}, {Vaillant, M.},
  {Valentini, G.}, {Valette, V.}, {van Elteren, A.}, {Van Hemelryck, E.}, {van
  Leeuwen, M.}, {Vaschetto, M.}, {Vecchiato, A.}, {Viala, Y.}, {Vicente, D.},
  {Vogt, S.}, {von Essen, C.}, {Voss, H.}, {Votruba, V.}, {Voutsinas, S.},
  {Walmsley, G.}, {Weiler, M.}, {Wertz, O.}, {Wevems, T.}, {Wyrzykowski, L.},
  {Yoldas, A.}, {Zerjal, M.}, {Ziaeepour, H.}, {Zorec, J.}, {Zschocke, S.},
  {Zucker, S.}, {Zurbach, C.}, \& {Zwitter, T.}}]{kinematics}
{Gaia Collaboration}, {Helmi, A.}, {van Leeuwen, F.}, {et~al.} 2018, A\&A, 616,
  A12.
\newblock \url{https://doi.org/10.1051/0004-6361/201832698}

\bibitem[{G{\'{o}}rski {et~al.}(2018)G{\'{o}}rski, Pietrzy{\'{n}}ski, Gieren,
  Graczyk, Suchomska, Karczmarek, Cohen, Zgirski, Wielg{\'{o}}rski, Pilecki,
  Taormina, Ko{\l}aczkowski, \& Narloch}]{Gorski_2018}
G{\'{o}}rski, M., Pietrzy{\'{n}}ski, G., Gieren, W., {et~al.} 2018, \apj, 156,
  278.
\newblock \url{https://doi.org/10.3847%2F1538-3881%2Faaeacb}

\bibitem[{{Harris}(2010)}]{H10}
{Harris}, W.~E. 2010, arXiv e-prints, arXiv:1012.3224

\bibitem[{{Hoyt} {et~al.}(2018){Hoyt}, {Freedman}, {Madore}, {Seibert},
  {Beaton}, {Hatt}, {Jang}, {Lee}, {Monson}, \& {Rich}}]{hoyt18}
{Hoyt}, T.~J., {Freedman}, W.~L., {Madore}, B.~F., {et~al.} 2018, \apj, 858, 12

\bibitem[{Hunter(2007)}]{Hunter:2007}
Hunter, J.~D. 2007, Computing In Science \& Engineering, 9, 90

\bibitem[{{Jang} {et~al.}(2020){Jang}, {Hoyt}, {Beaton}, {Freedman}, {Madore},
  {Lee}, {Neeley}, {Monson}, {Rich}, \& {Seibert}}]{maser_trgb}
{Jang}, I.~S., {Hoyt}, T., {Beaton}, R., {et~al.} 2020, arXiv e-prints,
  arXiv:2008.04181

\bibitem[{{Lee} {et~al.}(1993){Lee}, {Freedman}, \&
  {Madore}}]{lee_freedman_madore_1993}
{Lee}, M.~G., {Freedman}, W.~L., \& {Madore}, B.~F. 1993, \apj, 417, 553

\bibitem[{{Madore} \& {Freedman}(2020)}]{math_paper}
{Madore}, B.~F., \& {Freedman}, W.~L. 2020, \aj, 160, 170

\bibitem[{{Madore} {et~al.}(2018){Madore}, {Freedman}, {Hatt}, {Hoyt},
  {Monson}, {Beaton}, {Rich}, {Jang}, {Lee}, {Scowcroft}, \&
  {Seibert}}]{madore_2018_nirCalibration}
{Madore}, B.~F., {Freedman}, W.~L., {Hatt}, D., {et~al.} 2018, \apj, 858, 11

\bibitem[{{Pedregosa} {et~al.}(2012){Pedregosa}, {Varoquaux}, {Gramfort},
  {Michel}, {Thirion}, {Grisel}, {Blondel}, {M{\"u}ller}, {Nothman}, {Louppe},
  {Prettenhofer}, {Weiss}, {Dubourg}, {Vanderplas}, {Passos}, {Cournapeau},
  {Brucher}, {Perrot}, \& {Duchesnay}}]{sklearn}
{Pedregosa}, F., {Varoquaux}, G., {Gramfort}, A., {et~al.} 2012, arXiv
  e-prints, arXiv:1201.0490

\bibitem[{{Pietrzy{\'n}ski}(2019)}]{pietrzynski_2019}
{Pietrzy{\'n}ski}, G. 2019, \nat, 567, 200

\bibitem[{{Planck Collaboration} {et~al.}(2018){Planck Collaboration},
  {Aghanim}, {Akrami}, {Ashdown}, {Aumont}, {Baccigalupi}, {Ballardini},
  {Banday}, {Barreiro}, {Bartolo}, {Basak}, {Battye}, {Benabed}, {Bernard},
  {Bersanelli}, {Bielewicz}, {Bock}, {Bond}, {Borrill}, {Bouchet}, {Boulanger},
  {Bucher}, {Burigana}, {Butler}, {Calabrese}, {Cardoso}, {Carron},
  {Challinor}, {Chiang}, {Chluba}, {Colombo}, {Combet}, {Contreras}, {Crill},
  {Cuttaia}, {de Bernardis}, {de Zotti}, {Delabrouille}, {Delouis}, {Di
  Valentino}, {Diego}, {Dor{\'e}}, {Douspis}, {Ducout}, {Dupac}, {Dusini},
  {Efstathiou}, {Elsner}, {En{\ss}lin}, {Eriksen}, {Fantaye}, {Farhang},
  {Fergusson}, {Fernandez-Cobos}, {Finelli}, {Forastieri}, {Frailis},
  {Franceschi}, {Frolov}, {Galeotta}, {Galli}, {Ganga}, {G{\'e}nova-Santos},
  {Gerbino}, {Ghosh}, {Gonz{\'a}lez-Nuevo}, {G{\'o}rski}, {Gratton},
  {Gruppuso}, {Gudmundsson}, {Hamann}, {Handley}, {Herranz}, {Hivon}, {Huang},
  {Jaffe}, {Jones}, {Karakci}, {Keih{\"a}nen}, {Keskitalo}, {Kiiveri}, {Kim},
  {Kisner}, {Knox}, {Krachmalnicoff}, {Kunz}, {Kurki-Suonio}, {Lagache},
  {Lamarre}, {Lasenby}, {Lattanzi}, {Lawrence}, {Le Jeune}, {Lemos},
  {Lesgourgues}, {Levrier}, {Lewis}, {Liguori}, {Lilje}, {Lilley}, {Lindholm},
  {L{\'o}pez-Caniego}, {Lubin}, {Ma}, {Mac{\'{\i}}as-P{\'e}rez}, {Maggio},
  {Maino}, {Mandolesi}, {Mangilli}, {Marcos-Caballero}, {Maris}, {Martin},
  {Martinelli}, {Mart{\'{\i}}nez-Gonz{\'a}lez}, {Matarrese}, {Mauri}, {McEwen},
  {Meinhold}, {Melchiorri}, {Mennella}, {Migliaccio}, {Millea}, {Mitra},
  {Miville-Desch{\^e}nes}, {Molinari}, {Montier}, {Morgante}, {Moss}, {Natoli},
  {N{\o}rgaard-Nielsen}, {Pagano}, {Paoletti}, {Partridge}, {Patanchon},
  {Peiris}, {Perrotta}, {Pettorino}, {Piacentini}, {Polastri}, {Polenta},
  {Puget}, {Rachen}, {Reinecke}, {Remazeilles}, {Renzi}, {Rocha}, {Rosset},
  {Roudier}, {Rubi{\~n}o-Mart{\'{\i}}n}, {Ruiz-Granados}, {Salvati}, {Sandri},
  {Savelainen}, {Scott}, {Shellard}, {Sirignano}, {Sirri}, {Spencer},
  {Sunyaev}, {Suur-Uski}, {Tauber}, {Tavagnacco}, {Tenti}, {Toffolatti},
  {Tomasi}, {Trombetti}, {Valenziano}, {Valiviita}, {Van Tent}, {Vibert},
  {Vielva}, {Villa}, {Vittorio}, {Wandelt}, {Wehus}, {White}, {White},
  {Zacchei}, \& {Zonca}}]{planck_2018}
{Planck Collaboration}, {Aghanim}, N., {Akrami}, Y., {et~al.} 2018, ArXiv
  e-prints, arXiv:1807.06209

\bibitem[{{Price-Whelan} {et~al.}(2018){Price-Whelan}, {Sip{\H{o}}cz},
  {G{\"u}nther}, {Lim}, {Crawford}, {Conseil}, {Shupe}, {Craig}, {Dencheva},
  {Ginsburg}, {VanderPlas}, {Bradley}, {P{\'e}rez-Su{\'a}rez}, {de Val-Borro},
  {Paper Contributors}, {Aldcroft}, {Cruz}, {Robitaille}, {Tollerud},
  {Coordination Committee}, {Ardelean}, {Babej}, {Bach}, {Bachetti}, {Bakanov},
  {Bamford}, {Barentsen}, {Barmby}, {Baumbach}, {Berry}, {Biscani}, {Boquien},
  {Bostroem}, {Bouma}, {Brammer}, {Bray}, {Breytenbach}, {Buddelmeijer},
  {Burke}, {Calderone}, {Cano Rodr{\'\i}guez}, {Cara}, {Cardoso}, {Cheedella},
  {Copin}, {Corrales}, {Crichton}, {D{\textquoteright}Avella}, {Deil},
  {Depagne}, {Dietrich}, {Donath}, {Droettboom}, {Earl}, {Erben}, {Fabbro},
  {Ferreira}, {Finethy}, {Fox}, {Garrison}, {Gibbons}, {Goldstein}, {Gommers},
  {Greco}, {Greenfield}, {Groener}, {Grollier}, {Hagen}, {Hirst}, {Homeier},
  {Horton}, {Hosseinzadeh}, {Hu}, {Hunkeler}, {Ivezi{\'c}}, {Jain}, {Jenness},
  {Kanarek}, {Kendrew}, {Kern}, {Kerzendorf}, {Khvalko}, {King}, {Kirkby},
  {Kulkarni}, {Kumar}, {Lee}, {Lenz}, {Littlefair}, {Ma}, {Macleod},
  {Mastropietro}, {McCully}, {Montagnac}, {Morris}, {Mueller}, {Mumford},
  {Muna}, {Murphy}, {Nelson}, {Nguyen}, {Ninan}, {N{\"o}the}, {Ogaz}, {Oh},
  {Parejko}, {Parley}, {Pascual}, {Patil}, {Patil}, {Plunkett}, {Prochaska},
  {Rastogi}, {Reddy Janga}, {Sabater}, {Sakurikar}, {Seifert}, {Sherbert},
  {Sherwood-Taylor}, {Shih}, {Sick}, {Silbiger}, {Singanamalla}, {Singer},
  {Sladen}, {Sooley}, {Sornarajah}, {Streicher}, {Teuben}, {Thomas},
  {Tremblay}, {Turner}, {Terr{\'o}n}, {van Kerkwijk}, {de la Vega}, {Watkins},
  {Weaver}, {Whitmore}, {Woillez}, {Zabalza}, \& {Contributors}}]{astropy:2018}
{Price-Whelan}, A.~M., {Sip{\H{o}}cz}, B.~M., {G{\"u}nther}, H.~M., {et~al.}
  2018, \aj, 156, 123

\bibitem[{{Riess} {et~al.}(2019){Riess}, {Casertano}, {Yuan}, {Macri}, \&
  {Scolnic}}]{riess_2019}
{Riess}, A.~G., {Casertano}, S., {Yuan}, W., {Macri}, L.~M., \& {Scolnic}, D.
  2019, \apj, 876, 85

\bibitem[{{Rizzi} {et~al.}(2007){Rizzi}, {Tully}, {Makarov}, {Makarova},
  {Dolphin}, {Sakai}, \& {Shaya}}]{rizzi_2007}
{Rizzi}, L., {Tully}, R.~B., {Makarov}, D., {et~al.} 2007, \apj, 661, 815

\bibitem[{{Schlegel} {et~al.}(1998){Schlegel}, {Finkbeiner}, \&
  {Davis}}]{schlegel_1998}
{Schlegel}, D.~J., {Finkbeiner}, D.~P., \& {Davis}, M. 1998, \apj, 500, 525

\bibitem[{{Sch{\"o}nrich} {et~al.}(2019){Sch{\"o}nrich}, {McMillan}, \&
  {Eyer}}]{2019MNRAS.487.3568S}
{Sch{\"o}nrich}, R., {McMillan}, P., \& {Eyer}, L. 2019, \mnras, 487, 3568

\bibitem[{{Shapley} \& {Curtis}(1921)}]{shapley_curtis_1921}
{Shapley}, H., \& {Curtis}, H.~D. 1921, Bulletin of the National Research
  Council, 2, 171

\bibitem[{{Skrutskie} {et~al.}(2006){Skrutskie}, {Cutri}, {Stiening},
  {Weinberg}, {Schneider}, {Carpenter}, {Beichman}, {Capps}, {Chester},
  {Elias}, {Huchra}, {Liebert}, {Lonsdale}, {Monet}, {Price}, {Seitzer},
  {Jarrett}, {Kirkpatrick}, {Gizis}, {Howard}, {Evans}, {Fowler}, {Fullmer},
  {Hurt}, {Light}, {Kopan}, {Marsh}, {McCallon}, {Tam}, {Van Dyk}, \&
  {Wheelock}}]{2mass_2006}
{Skrutskie}, M.~F., {Cutri}, R.~M., {Stiening}, R., {et~al.} 2006, \aj, 131,
  1163

\bibitem[{Stetson(1987)}]{Stetson_1987}
Stetson, P.~B. 1987, Publications of the Astronomical Society of the Pacific,
  99, 191.
\newblock \url{https://doi.org/10.1086%2F131977}

\bibitem[{{Stetson}(1992)}]{allstar_1992}
{Stetson}, P.~B. 1992, in Astronomical Society of the Pacific Conference
  Series, Vol.~25, Astronomical Data Analysis Software and Systems I, ed. D.~M.
  {Worrall}, C.~{Biemesderfer}, \& J.~{Barnes}, 297

\bibitem[{{Stetson}(1994)}]{allframe}
{Stetson}, P.~B. 1994, \pasp, 106, 250

\bibitem[{Stetson {et~al.}(2019)Stetson, Pancino, Zocchi, Sanna, \&
  Monelli}]{S19}
Stetson, P.~B., Pancino, E., Zocchi, A., Sanna, N., \& Monelli, M. 2019,
  Monthly Notices of the Royal Astronomical Society, 485, 3042.
\newblock \url{https://doi.org/10.1093/mnras/stz585}

\bibitem[{{Thompson} {et~al.}(2001){Thompson}, {Kaluzny}, {Pych}, {Burley},
  {Krzeminski}, {Paczy{\'n}ski}, {Persson}, \& {Preston}}]{thompson_2001}
{Thompson}, I.~B., {Kaluzny}, J., {Pych}, W., {et~al.} 2001, \aj, 121, 3089

\bibitem[{{Thompson} {et~al.}(2020){Thompson}, {Udalski}, {Dotter}, {Rozyczka},
  {Schwarzenberg-Czerny}, {Pych}, {Beletsky}, {Burley}, {Marshall},
  {McWilliam}, {Morrell}, {Osip}, {Monson}, {Persson}, {Szyma{\'n}ski},
  {Soszy{\'n}ski}, {Poleski}, {Ulaczyk}, {Wyrzykowski}, {Koz{\l}owski},
  {Mr{\'o}z}, {Pietrukowicz}, \& {Skowron}}]{thompson_2020}
{Thompson}, I.~B., {Udalski}, A., {Dotter}, A., {et~al.} 2020, \mnras, 492,
  4254

\bibitem[{{Van Der Walt} {et~al.}(2011){Van Der Walt}, {Colbert}, \&
  {Varoquaux}}]{numpy:2011}
{Van Der Walt}, S., {Colbert}, S.~C., \& {Varoquaux}, G. 2011, Computing in
  Science \& Engineering, 13, 22

\bibitem[{{Vasiliev}(2019)}]{V19}
{Vasiliev}, E. 2019, \mnras, 484, 2832

\end{thebibliography}

\end{document}